\documentclass[11pt,a4paper]{article}
\pdfoutput=1
\usepackage[utf8]{inputenc}
\usepackage[T1]{fontenc}
\usepackage[pdftex]{graphicx}
\usepackage{lmodern}
\usepackage{float}
\usepackage{amssymb,amsmath,amsthm,amsfonts,mathtools,mathrsfs}
\usepackage{hyperref}
\usepackage{booktabs}

\usepackage{microtype}
\usepackage[usenames,dvipsnames,svgnames,table]{xcolor}
\usepackage[left=2.5cm,right=2.5cm,bottom=3cm,top=3cm]{geometry}

\usepackage{braket}
\usepackage{xspace}
\usepackage{enumerate}
\usepackage{caption}
\usepackage{pstricks}         

\usepackage[font=small,labelfont=bf]{caption}
\usepackage{array}
\usepackage{cancel}
\usepackage{wrapfig}
\usepackage{tabularx}
\usepackage{colortbl}
\usepackage[usenames,dvipsnames]{xcolor}
\usepackage{arydshln}

\hypersetup{
	colorlinks=true,
	linktoc=page,
	citecolor=RedViolet,
	linkcolor=RedViolet,
	urlcolor=Black} 

\numberwithin{equation}{section}

\makeatletter
\renewcommand{\maketitle}
{ \begingroup \begin{center} \large {\bf \@title}
		\vskip 5pt \large \@author \\ \vskip 5pt \@date \end{center}
	\vskip 5pt \endgroup \setcounter{footnote}{0} }
\makeatother 

\usepackage{tikz}
\usetikzlibrary{matrix,arrows,shapes}
\usetikzlibrary{positioning}
\usetikzlibrary{fadings}
\usetikzlibrary{decorations.markings}

\def\vacuumheight{1}

\def\labelvdist{0.3}

\def\labelddist{\labelvdist*0.70710678118}

\definecolor{grayn}{HTML}{D4D4D4}

\newlength{\vacuumradius}
\newlength{\onshellradius}
\tikzstyle{db}=[circle, black, fill=black, minimum width=\onshellradius, draw, inner sep=0pt]
\tikzstyle{dot}=[circle, black, fill=black, minimum width=0.1cm, draw, inner sep=0pt]
\tikzstyle{dw}=[circle, black, fill=white, minimum width=\onshellradius, draw, inner sep=0pt]
\tikzstyle{dvac}=[circle, black, fill=lightgrayn, minimum width=\vacuumradius, inner sep=0pt]
\tikzstyle{dl}=[circle, black, fill=white, inner sep=2pt]

\newcommand{\drawminimalff}[1]{
	\draw[thick,double] (#1-0.5,-0) -- (#1-0.5,-0.5); 
	\draw (#1-0.5,-0.5) -- (#1-1,-\vacuumheight);  
	\draw (#1-0.5,-0.5) -- (#1,-\vacuumheight);}

\newcommand{\comments}[1]{}
\newcommand{\la}{\langle}

\newcommand{\F}{\mathcal{F}}
\newcommand{\T}{\mathcal{T}}

\newcommand{\N}{\mathcal{N}}

\newcommand{\Tr}{\text{Tr}}

\renewcommand{\b}[1]{\braket{#1}}
\newcommand{\LL}{\mathcal{L}}
\renewcommand{\O}{\mathcal{O}}
\newcommand{\pole}[1]{(#1)}
\newcommand{\poleu}[1]{(\underline{{#1}})}
\newcommand{\contour}{\mathcal{C}^{\mathrm{BCFW}}}
\newcommand{\eqndot}{\; . }
\newcommand{\eqncom}{\; , }

\newcommand{\vll}{{\smash{\lambda}}}
\newcommand{\vlt}{{\smash{\tilde{\lambda}}}}
\newcommand{\vle}{{\smash{\tilde{\eta}}}}
\newcommand{\abr}[1]{\langle #1 \rangle}
\newcommand{\sbr}[1]{\left[ #1 \right]}
\newcommand{\splus}{\! + \!}
\newcommand{\sminus}{\! - \!}
\newcommand{\ssep}{\;}

\newcommand{\dd}{\ensuremath{\mathrm{d}\xspace}}
\newcommand{\amp}{\mathsf{A}}
\newcommand{\ff}{\mathsf{F}}
\newcommand{\gi}{\mathcal{G}}

\newcommand{\calN}{\mathcal{N}}
\newcommand{\nfsym}{\ensuremath{\calN\!\!=\!4} SYM\xspace}

\title{A note on NMHV form factors from the\\ Graßmannian and the twistor string}
\author{
	David Meidinger,
	Dhritiman Nandan,
	Brenda Penante,
	Congkao Wen
}

\begin{document}
	
	\begingroup\parindent0pt
	\begin{flushright}\footnotesize
		\texttt{HU-MATH-2017-05}\\
		\texttt{HU-EP-17/16}\\
		\texttt{CERN-TH-2017-139}\\
			\texttt{Edinburgh 2017/14}
	\end{flushright}
	\let\centering\relax
	\vspace*{3em}
	\begingroup\huge
	\centering
	\bf
	\makeatletter\@title\makeatother
	\par\endgroup
	\vspace{3.5em}
	\begingroup\large
	\centering
	{\bf David Meidinger}$\,^a$, 
	{\bf Dhritiman Nandan}$\,^{b,c}$, 
	{\bf Brenda Penante}$\,^d$, 
	{\bf Congkao Wen}$\,^{e,f,c}$
	\par\endgroup
	\vspace{2em}
	\begingroup\sffamily\footnotesize
	\centering
	\begin{tabular}{l@{\hskip 3pt}l}
		$a$ &Institut für Mathematik und Institut für Physik,
		Humboldt-Universität zu Berlin,\\
		& IRIS Gebäude,
		Zum Großen Windkanal 6,
		12489 Berlin, Germany\\
		& david.meidinger@physik.hu-berlin.de\\
		&\\
		$b$ & Higgs Centre for Theoretical Physics, School of Physics and Astronomy,\\
		& The University of Edinburgh, Edinburgh EH9 3JZ, Scotland, UK\\
		& dhritiman.nandan@ed.ac.uk\\
		&\\
		$c$ & Kavli Institute for Theoretical Physics\\
		& University of California, Santa Barbara, CA 93106.\\
		&\\
		$d$ & CERN Theory Division\\
		& 1211 Geneva 23, Switzerland\\
		& b.penante@cern.ch\\
		&\\
		$e$   &Walter Burke Institute for Theoretical Physics\\
		& California Institute of Technology, Pasadena, CA 91125, U.S.A.  \\
		& cwen@caltech.edu\\
		&\\
		$f$ & Mani L. Bhaumik Institute for Theoretical Physics\\
		& Department of Physics and Astronomy, UCLA, Los Angeles, CA 90095, U.S.A.\\

	\end{tabular}
	\par\endgroup
	
	\vspace{1.5cm}
	\noindent
	\textbf{Abstract}\\
	In this note we investigate Graßmannian formulas for form factors 
	of the chiral part of the stress-tensor multiplet in $\mathcal{N}\!=\!4$ superconformal 
	Yang-Mills theory. 
	We present an all-$n$ contour for the $G(3,n+2)$ Gra\ss mannian integral of NMHV form factors derived from on-shell diagrams
	and the BCFW recursion relation. In addition, we study other $G(3,n+2)$ formulas obtained from the connected prescription introduced recently.
	We find a recursive expression for all $n$ and study its properties. 
	For $n\geq 6$, our formula has the same recursive structure as its amplitude
	counterpart, making its soft behaviour manifest. 
	Finally, we explore the connection between the two Graßmannian formulations, using the global residue theorem, and find that it is much more intricate compared to scattering amplitudes.
	
    \vspace{3\baselineskip}
	\endgroup
	
	\thispagestyle{empty}
	
	\newpage
	\tableofcontents
	\vspace{2\baselineskip}
	
	\section{Introduction}
	
	Although the study of analytic properties of scattering amplitudes in general field theories is an old subject in Physics, no theory has seen a rate of development as steep as maximally supersymmetric Yang-Mills theory ($\N=4$ SYM) in the planar limit. Scattering amplitudes in $\N=4$ SYM became a subject of intense study in particular after a duality with a topological twistor string theory was proposed in \cite{Witten:GaugeAsStringInTwistor2003}. This sparked a tremendous amount of work which, among other results,
	allowed the hidden symmetries and the integrability \cite{Beisert:2010jr} of the theory to become apparent from the perspective of scattering amplitudes \cite{Drummond:2009fd,Ferro:2013dga,Chicherin:2013ora}.
	A key role in these developments was played by novel formulations of scattering
	amplitudes.
	Among the various streams of results in this regard is a representation of tree-level scattering amplitudes and loop level leading singularities as contour integrals over a Graßmannian space \cite{ArkaniHamed:2009dn}. This representation led to the emergence of the on-shell diagram formalism \cite{ArkaniHamed:2012nw} and finally to the amplituhedron \cite{Arkani-Hamed:2013jha,Arkani-Hamed:2013kca},
	providing a new, geometrical perspective on amplitudes, hidden in the usual
	space-time formulation.\footnote{See also \cite{Eden:2017fow} for a geometric picture of correlation functions in $\mathcal{N}=4$ SYM.}
	
	A natural question one may ask is whether similar geometrical formulations hold for quantities which are more generic than on-shell scattering amplitudes, for instance form factors involving off-shell gauge invariant operators $\O(x)$, defined as the matrix element of an operator taken between the vacuum and an on-shell state of $n$ particles,
	\begin{align}
	\begin{split}
	\mathcal{F}_{\O}(1,\dots,n;q)\equiv &\int\! \dd^4x\, e^{-iq x}\la 1\ldots n|\mathcal{O}(x)|0\rangle\ .
	\end{split}
	\end{align}
	In addition to the on-shell momenta $p_i,\,i=1,\dots,n$ satisfying $p_i^2=0$, a form factor depends on the momentum $q$ conjugate to the position of the operator. This momentum, unlike those of the on-shell particles, is in general not light-like
	
	The operator with the most well-studied form factors \cite{Brandhuber:2010ad,Brandhuber:2011tv,Bork:2012tt,Bork:2014eqa,Bork:2016hst,Bork:2017qyh} is the chiral part of the stress-tensor multiplet $\T(x,\theta^+)$, which is a protected supersymmetric operator. It can be  expanded in harmonic superspace%
\footnote{We follow closely the notation and conventions of 
		\cite{Eden:2011yp,Eden:2011ku} for the harmonic projections, see also \cite{Brandhuber:2011tv}. Note that since we are studying the chiral part of the stress tensor multiplet the $ \theta^- $ is set to zero in this notation.
        The fermionic variables associated to the on-shell particles will be denoted as $\eta^{+a}$
        and $\eta^{-a}$.
    }
    Graßmann coordinates $\theta_\alpha^{+ a}$, with $\alpha, a = 1,2$, and contains the operator $\Tr(\phi^2)$, with $\phi$ one of the scalars of the theory, as the top component and the on-shell Lagrangian of \nfsym as the coefficient of the highest power in $\theta^+$.
	Form factors of operators belonging to the same supersymmetric multiplet can be combined into a supersymmetric form factor as
	\begin{align}
	\F_{\T}(1,\dots,n;q,\gamma^-) = \int\! \dd^4 x \,\dd^4 \theta^+ e^{-iq x -i\theta^{+a}_\alpha \gamma_a^{-\alpha}} \langle 1,\dots,n|\T(x,\theta^+)|0\rangle\eqncom
	\end{align}
	where $\gamma^{- \alpha}_{a}$ is the variable conjugate to the superspace coordinate $\theta_\alpha^{+a}$. Like scattering amplitudes, this expression admits an expansion in MHV degrees, $ k $. In the following we denote by $\ff_{n,k}$ the colour ordered N$^{k-2}$MHV form factor of $\T$ with $n$ on-shell states, and by $\amp_{n,k}$ its amplitude counterpart.
	
	In the Gra\ss mannian formulation of \cite{ArkaniHamed:2009dn}, $\amp_{n,k}$ is represented as a contour integral over the Gra\ss mannian $G(k,n)$, which is the space of $k$-dimensional planes in $\mathbb{C}^n$. 
	In \cite{Frassek:2015rka} on-shell diagrams and an associated Graßmannian formula were presented for tree-level form factors of the operator $\T$, using a parametrization of the operator momentum as a sum of two on-shell momenta.
	From the Graßmannian integral, N$^{k-2}$MHV form factors of $\T$ can be obtained from a combination of residues in $G(k,n+2)$ \cite{Frassek:2015rka}. Compared to scattering amplitudes, some difficulties arise as a result of the operator being a colour singlet and not participating in the colour ordering of the external particles: for instance, there exist $n$ cyclically related top forms on the Graßmannian, and no single form contains
	all residues which build up the tree level form factor. As a result, residues from different top forms must be combined in a way that was, until now, only known on a case by case basis.
	
In this note we address this matter, providing a general contour prescription for the Graßmannian formulation of \cite{Frassek:2015rka}. To this end we utilise the correspondence between cells of the Graßmannian and on-shell diagrams. 
 We find a recursive solution to the Britto-Cachazo-Feng-Witten (BCFW) recursion relation \cite{Britto:2005fq}, ensuring that the residues reproduce all factorization poles. This yields the analogue of the tree-level contour defined in \cite{ArkaniHamed:2009dn} for scattering amplitudes.
While we focus on NMHV form factors, this technique can be applied for general MHV degree.
 A corollary of our finding is that no linear combination of top forms can be taken to reproduce the form factor when endowed with a joint contour prescription for all forms. Rather, individual residues from different top forms must be picked individually, but nevertheless systematically.\footnote{This was already observed for amplitudes in $\N=8$ supergravity beyond the MHV case \cite{Farrow:2017eol}.} We also discuss ambiguities concerning the choice of top form for each residue.
	
	The Graßmannian formulation of scattering amplitudes was shown to be tightly related to the twistor string theory formalism \cite{Spradlin:2009qr, Dolan:2009wf}. The connection between these two approaches was realised by expressing the Roiban-Spradlin-Volovich (RSV) formulas for tree-level amplitudes in $\mathcal{N}=4$ SYM \cite{Roiban:2004yf} in terms of the link variables introduced in  \cite{ArkaniHamed:2009si}. The kinematic constraints in the RSV picture do not leave any free integration variables, and recasting the formulas as integrals over the Graßmannian \cite{Nandan:2009cc} via the link variables played an important role in the formulation of the general tree-level contour for the amplitude Graßmannian integral \cite{ArkaniHamed:2009dg,Bourjaily:2010kw}.
	A generalization of the RSV prescription for form factors was developed in \cite{Brandhuber:2016xue} and \cite{He:2016jdg}. In particular, \cite{Brandhuber:2016xue} put forward a link representation. In this work, we perform the last step into lifting the link representation to the Gra\ss mannian. This provides a different Graßmannian representation compared to the integral obtained from on-shell diagrams, with a fixed contour of integration. We find a recursive definition of the formula and show that it can be interpreted as the ``inverse soft'' addition of particles, identical to the structure of amplitudes~\cite{ArkaniHamed:2009dg}.
	
	For scattering amplitudes, the fact that the two Graßmannian formulations---based on on-shell diagrams and on the connected prescription---lead to the same result can be shown through successive applications of the global residue theorem (GRT) \cite{Nandan:2009cc}. In addition, it is possible to define a family of Graßmannian formulas parametrised by a smooth parameter $t$ such that it returns the connected prescription for $t=1$ and the standard $G(k,n)$ formula endowed with the tree level contour for $t=0$ or $t=\infty$. In this work we investigate similar relations between the two Graßmannian formulas for NMHV form factors. In particular, we show that a smooth deformation between them is not available, although the application of successive GRTs can uncover the BCFW poles from the connected formula for four and five points. Starting from six points, the relation between the two representations becomes very subtle; we show that the Graßmannian integral from the connected prescription does not possess all BCFW factorization poles in a way accessible via the GRT.
	
	This note is organised as follows. In Section~\ref{sec:contour} we study the BCFW recursion relation in terms of on-shell diagrams and derive a compact formula for the form-factor Graßmannian contour in the NMHV case. In Section~\ref{sec:connected} we lift the link representation of \cite{Brandhuber:2016xue} to a second Graßmannian formula for NMHV form factors. We study different representations of this formula and show that it can be written in a way which closely mirrors the corresponding representation of amplitudes. Section~\ref{sec:relation} is devoted to relating the two Graßmannian formulas for NMHV form factors by means of the GRT.
	
	\section{The NMHV contour for the form factor Graßmannian}
	\label{sec:contour}
	
	The Gra\ss mannian formulation for N$^{k-2}$MHV form factors was introduced in \cite{Frassek:2015rka}, where a form factor top form in $G(k,n+2)$ was first written down. This formulation lacked a contour prescription, and the combination of residues that compose a given form factor---originating in general from different top forms related by cyclic symmetry---was worked out case by case. In this section, we present a closed formula for the tree-level contour for NMHV form factors. This provides a systematic way of computing form factors of the chiral part of the stress-tensor operator for any $n$.
	
	\subsection{Brief review of the Graßmannian integral for NMHV form factors}
	\label{sec:Grassmannian}
	
	In \cite{Frassek:2015rka} it was shown that form factors
	of the chiral stress-tensor multiplet in \nfsym can be represented
	via a generalization of on-shell diagrams \cite{ArkaniHamed:2012nw}.
	These diagrams use the minimal, i.e. two point form factor as 
	a vertex, in addition to the two three-point amplitudes, 
	\begin{equation}
	\label{eq:vertices}
	\begin{aligned}
	\begin{aligned}
	\begin{tikzpicture}[scale=0.58]
	\draw (1,1) -- (1,1+0.65);
	\draw (1,1) -- (1-0.5,1-0.5);
	\draw (1,1) -- (1+0.5,1-0.5);
	\node [] at (1,1+0.65+\labelvdist) {\footnotesize$1$};
	\node [] at (1-0.5-0.2,1-0.5-0.2) {\footnotesize$3$};
	\node [] at (1+0.5+0.2,1-0.5-0.2) {\footnotesize$2$};
	\node[db] at (1,1) {};
	\end{tikzpicture}
	\end{aligned}
	&=
	\amp_{3,2}(1,2,3)=\frac{
		\delta^4(\vll_1\vlt_1+\vll_2\vlt_2+\vll_3\vlt_3)
		\delta^8(\vll_1\vle_1+\vll_2\vle_2+\vll_3\vle_3)
	}{\abr{12}\abr{23}\abr{31}} \eqncom\\
	\begin{aligned}
	\begin{tikzpicture}[scale=0.58]
	\draw (1,1) -- (1,1+0.65);
	\draw (1,1) -- (1-0.5,1-0.5);
	\draw (1,1) -- (1+0.5,1-0.5);
	\node [] at (1,1+0.65+\labelvdist) {\footnotesize$1$};
	\node [] at (1-0.5-0.2,1-0.5-0.2) {\footnotesize$3$};
	\node [] at (1+0.5+0.2,1-0.5-0.2) {\footnotesize$2$};
	\node[dw] at (1,1) {};
	\end{tikzpicture}
	\end{aligned}
	&=
	\amp_{3,1}(1,2,3)
	=\frac{
		\delta^4(\vll_1\vlt_1+\vll_2\vlt_2+\vll_3\vlt_3)
		\delta^4(\sbr{12}\vle_3+\sbr{23}\vle_1+\sbr{31}\vle_2)
	}{\sbr{12}\sbr{23}\sbr{31}} \eqncom
	\\
	\begin{aligned}
	\begin{tikzpicture}[scale=0.58]
	\drawminimalff{1} 
	\node [] at (-\labelddist,-\vacuumheight-\labelddist) {\footnotesize$2$};
	\node [] at (1+\labelddist,-\vacuumheight-\labelddist) {\footnotesize$1$};
	\end{tikzpicture}
	\end{aligned}
	&=
	\ff_{2,2}(1,2;q,\gamma^-)=\frac{
		\delta^4(\vll_1\vlt_1 + \vll_2\vlt_2 - q)
		\delta^4(\vll_1\vle_1^+ + \vll_2\vle_2^+ )
		\delta^4(\vll_1\vle_1^- + \vll_2\vle_2^- - \gamma^-)
	}{\abr{12}\abr{21}} \eqndot
	\end{aligned}
	\end{equation}
	Generic form-factor on-shell diagrams are obtained by gluing the fundamental vertices above, i.e. by performing an integration over the one-particle on-shell phase space for each internal edge.
	The parametrization of the off-shell momentum $q$ and supermomentum $\gamma^-$ is done via the addition of two auxiliary on-shell particles. We label these particles by $x$ and $y$ in order to distinguish them from the $n$ on-shell states of the form factor. Concretely, let $\vll_x$ and $\vll_y$ be arbitrary (non-collinear) reference spinors, and define
	\begin{equation}
	\begin{aligned}
	\vlt_{x}&=-\frac{\bra{y}q}{\abr{yx}}\eqncom \quad &
	\vle_{x}^-&=-\frac{\bra{y}\gamma^-}{\abr{yx}}\eqncom\quad &
	\vle_x^+&=0\ ,\\
	\vlt_{y}&=-\frac{\bra{x}q}{\abr{xy}}\eqncom &
	\vle_{y}^-&=-\frac{\bra{x}\gamma^-}{\abr{xy}}\eqncom &
	\vle_y^+&=0\ ,
	\end{aligned}
	\label{eq:kinematics}
	\end{equation}
	such that 
	$\vll_x\vlt_x+\vll_y\vlt_y=p_x+p_y=-q$, 
	$\vll_x\vle^-_x+\vll_y\vle^-_y=-\gamma^-$ and $\vll_x\vle^+_x+\vll_y\vle^+_y=0$.
	
	\bigskip\noindent
	Using these variables, the Graßmannian 
	formula for NMHV form factors is given by \cite{Frassek:2015rka}
	\begin{equation}
	\gi_{n,3}^{[s]} = 
	\abr{xy}^2
	\int\frac{\dd^{3\times(n+2)}C}{\text{Vol}[GL(3)]}\;
	\frac{
		\delta^{2\times 3}(C\cdot\vlt) \, \delta^{4\times 3}(C\cdot\vle) \, \delta^{2\times(n-1)}(C^\perp\cdot\vll)
	}{     \big[
	\pole{1}\cdots\pole{n-2} \;\; \poleu{1}\poleu{n} \;\; (xy\,(n\sminus1\ssep n)\cap(12))
	\big]}_{\smash{\sigma_s}
}\ .
\label{eq:gi}
\end{equation}
The notation used here is as follows. $C$ is a $3\times(n+2)$ matrix parametrizing $G(3,n+2)$,
\begin{equation} \label{eq:vectorC}
\begin{pmatrix}
C_{1}, & C_2, & \cdots, & C_{n-1}, & C_{n}, & C_x, & C_y     \end{pmatrix} \, ,
\end{equation}
where each column $C_i$ is a $k$-dimensional vector, namely $C^{\rm T}_i\equiv (C_{1i} \, , C_{2i} \, , \cdots, C_{ki})$ (for NMHV $k=3$). 
We abbreviate minors of $C$ which are consecutive in the $n$ labels corresponding to the on-shell particles with a single label, as in %
\cite{ArkaniHamed:2009dg,Bourjaily:2010kw}, 
and use a similar notation for minors involving the columns with labels
$x$ and $y$,
\begin{equation}
\label{eq:poles}
\pole{i} \equiv (i\ssep i\splus1\ssep i\splus2)
\eqncom
\qquad
\poleu{i} \equiv (i \ssep x\ssep y)
\eqndot
\end{equation}
Furthermore, we employ the standard notation 
$(ij)\cap(kl) \equiv C_i (jkl) - C_j (ikl)$
for the intersection of the lines $(ij)$ and $(kl)$.\footnote{We observe that the occurrence of poles of the form $ (xy\,(a b)\cap(cd)) $  in \eqref{eq:gi} is similar to those found in  \cite{Franco:2015rma,Bourjaily:2016mnp} for non-planar on-shell diagrams.}
Finally,
$\sigma_s$ is a cyclic shift of the on-shell labels by $s$
appearing in the integrand,
\begin{equation}
\sigma_s = \begin{pmatrix}
1 & 2 & \cdots & n-1 & n & x & y \\
\downarrow &
\downarrow &
&
\downarrow &
\downarrow &
\downarrow &
\downarrow \\
1+s & 2+s & & n-1+s & n+s & x & y
\end{pmatrix}
\quad \text{with } i+n \simeq i\eqncom
\end{equation}
reflecting the fact that the insertion of the colourless operator in the on-shell diagram artificially breaks the cyclic invariance in the on-shell labels. This leads to $n$ inequivalent
top forms labelled by the shift $s$.
The Graßmannian integral \eqref{eq:gi} is the form factor analogue of the NMHV amplitude formula \cite{ArkaniHamed:2009dn}
\begin{equation}
\label{eq:ampG}
\LL^{\rm amp}_{n,3} = \int_{\Gamma^{\rm BCFW}_{n,3}} \frac{\dd^{3\times n}C}{\text{Vol}[GL(3)]}\;
\frac{
	\delta^{2\times 3}(C\cdot\vlt) \, \delta^{4\times 3}(C\cdot\vle) \, \delta^{2\times(n-3)}(C^\perp\cdot\vll)
}{
\pole{1} \pole{2} \cdots\pole{n}
}\eqncom
\end{equation}
which is equipped with the BCFW contour $\Gamma^{\rm BCFW}_{n,3}$, whose general expression is known \cite{ArkaniHamed:2009dg}. For an $n$-point amplitude, there are $(n-5)$ free integration variables
$ \tau_1,\ldots, \tau_{n-5} $. We employ the following notation for the residues:
\begin{equation}
\{f_1,f_2,\dots,f_{n-5}\}\quad \leftrightarrow\quad \text{Residue of Gra\ss mannian integral around poles $|\tau_i - f_i| = \epsilon_i \rightarrow 0$\ . }
\label{residuenotation}
\end{equation}
The tree-level contour can then be specified by $(n-5)$ vanishing minors $\{(i_1) \, ,(i_2) \, ,\cdots, (i_{n-5}) \}$.
Using this notation, the NMHV BCFW contour takes an ``odd-even'' pattern,  explicitly given by 
\begin{equation}
\label{eq:ampContour}
\Gamma^{\rm BCFW}_{n,3} = \mathscr{O} \star  \mathscr{E}  \star  \mathscr{O}  \star  \mathscr{E}   \star 
\cdots \, ,
\end{equation}
where $\mathscr{O}$ is the set of odd numbered particles and $\mathscr{E}$ is the set of even numbered particles, 
\begin{equation}
\mathscr{O} = \sum_{i \in {\rm Odd}} \{(i) \} \, , \quad 
\mathscr{E} = \sum_{i \in {\rm Even}} \{(i) \} \,,
\end{equation}
and the product $\star$ is defined as
\begin{equation}
\{ (i)\} \star \{ (j) \}  =     \left\{ \begin{array}{rcl}
&      \{ (i) , (j) \}  & \mbox{for}
\quad i<j
\\
\\
& 0 & \mbox{for}  \quad i>j
\end{array}\right. \eqndot
\end{equation}
The aim of this section is to present a similar closed formula for the tree-level contour for NMHV form factors. Unlike amplitudes, in principle top forms with different values of shift parameter $s$ must be combined together in order to reproduce all factorization poles of the form factor.

\subsection{Closed form of the contour}
\label{sec:recursion}

In this section, we derive a closed formula for the NMHV tree-level contour for the form factor formula \eqref{eq:gi} from the BCFW recursion relation. Due to the fact that multiple top forms have to be considered,
it turns out that the contour cannot be thought of as a single domain of integration,
but rather as a set of contours for the individual top forms.
We express it  as a list of poles which are in one-to-one correspondence with the BCFW terms, 
the residues of which add up to the tree-level form factor. These residues may come from distinct top forms (different values of the shift $s$), but we argue that every choice of $s$ produces the same residue, provided the corresponding form has a non-vanishing residue on the respective
configuration.

After solving the kinematical constraints, the NMHV form factors with $n$ legs is a contour integral in $n-3$ variables $\tau_1,\dots,\tau_{n-3}$, expressed as $ \{f_1,f_2,\dots,f_{n-3}\}$ following the notation \eqref{residuenotation}.
Using \eqref{eq:poles}, the tree-level NMHV $n$-point form factor is now given by the combination of residues
\begin{equation}
\label{eq:FFcontour}
\contour_{n,3} = \sum_{m=0}^{n-3}
\left[
R_m^{00}
+ \sum_{i_1=1}^m  R^{i_10}_m
+ \sum_{i_2=1}^{n-m-3}  R^{0i_2}_m
\right]\ ,
\end{equation}
where each residue above reads
\begin{equation} 
R^{i_1 i_2}_m \coloneqq
\Bigg\{
\underbrace{
	\underbrace{
		\poleu{1},\ldots,\poleu{i_1}
	}_{i_1}
	,
	\pole{i_1+1},\ldots,\pole{m}
}_m
,
\underbrace{
	\underbrace{
		\poleu{m+3},\ldots,\poleu{m+i_2+2}
	}_{i_2}
	,
	\pole{m+i_2+3},\ldots,\pole{n-1}
}_{n-m-3}
\Bigg\}\eqndot
\end{equation}
Note that \eqref{eq:FFcontour} makes no mention of the shift $s$ that labels the top form in \eqref{eq:gi}. The reason is that for each term, one can take the residue from any top form (using any shift), as long as this form has a pole at the desired configuration. As we show shortly, each term in  \eqref{eq:FFcontour} corresponds to a particular BCFW factorization, and the degeneracy in $s$ follows from the cyclic symmetry of a sub-form factor entering the recursion relation.  To be explicit, we can summarise the possible choices for $s$:
\begin{equation*}
\begin{tabular}{lccccc}
\toprule
terms:\qquad\quad &
$R^{i_1 0}_m$ &
$R^{0 i_2}_m$  &
$R^{0 0}_0$ &
$R^{0 0}_{n-3}$&   
$R^{0 0}_{m}$
\\
\midrule
shifts  $s$: \qquad\quad&
$0,1,\ldots,i_1$ \quad&
$m+2,\ldots,m+2+i_2$ \quad&
$1,2$ \quad&
$0,n-1$ \quad&
$m+2$\\
\bottomrule
\end{tabular}
\end{equation*}
The closed formula for the contour  \eqref{eq:FFcontour} follows from the BCFW recursion relation \cite{Britto:2005fq}, which can be depicted graphically for NMHV form factors as \cite{Brandhuber:2010ad,Brandhuber:2011tv}
\begin{equation}
\label{eq:bcfwnmhv}
\ff_{n,3}=
\sum_{n_l = 2}^{n-2}
\!
\begin{aligned}
\begin{tikzpicture}[scale=0.7]
\draw (0,-0) -- (2,-0); 
\draw (0,-1.5) -- (2,-1.5); 
\draw (0,-0) -- (0,-2.25); 
\draw (2,-0) -- (2,-2.25); 
\draw (0,-0) -- (-1.2,-0);
\draw (0,-0) -- (-0,+1.2);
\draw (2,-0) -- (+3.2,-0);
\draw (2,-0) -- (+2,+1.2);
\node[] at (-0.7,+0.7) {\rotatebox{45}{$\cdots$}};
\node[] at (+2.7,+0.7) {\rotatebox{-45}{$\cdots$}};
\node[dw] at (0,-1.5) {};
\node[db] at (2,-1.5) {};
\draw [thick,double] (0,0) -- (-1,-1);
\node[circle, black, fill=grayn, minimum width=5*\onshellradius, draw, inner sep=0pt] at (0,0) {$\scriptstyle \ff_{n_l,2}$};
\node[circle, black, fill=grayn, minimum width=5*\onshellradius, draw, inner sep=0pt] at (2,0) {$\scriptstyle \amp_{n_r,2}$};
\node[] at (0,-2.25-\labelvdist) {\footnotesize$1$};
\node[] at (2,-2.25-\labelvdist) {\footnotesize$n$};
\end{tikzpicture}
\end{aligned}
+\;
\sum_{n_l = 3}^{n}
\!\!\!
\begin{aligned}
\begin{tikzpicture}[scale=0.7]
\draw (0,-0) -- (2,-0); 
\draw (0,-1.5) -- (2,-1.5); 
\draw (0,-0) -- (0,-2.25); 
\draw (2,-0) -- (2,-2.25); 
\draw (0,-0) -- (-1.2,-0);
\draw (0,-0) -- (-0,+1.2);
\draw (2,-0) -- (+3.2,-0);
\draw (2,-0) -- (+2,+1.2);
\node[] at (-0.7,+0.7) {\rotatebox{45}{$\cdots$}};
\node[] at (+2.7,+0.7) {\rotatebox{-45}{$\cdots$}};
\node[dw] at (0,-1.5) {};
\node[db] at (2,-1.5) {};
\draw [thick,double] (2,0) -- (+3,-1);
\node[circle, black, fill=grayn, minimum width=5*\onshellradius, draw, inner sep=0pt] at (0,0) {$\scriptstyle \amp_{n_l,2}$};
\node[circle, black, fill=grayn, minimum width=5*\onshellradius, draw, inner sep=0pt] at (2,0) {$\scriptstyle \ff_{n_r,2}$};
\node[] at (0,-2.25-\labelvdist) {\footnotesize$1$};
\node[] at (2,-2.25-\labelvdist) {\footnotesize$n$};
\end{tikzpicture}
\end{aligned}
\;
+ 
\begin{aligned}
\begin{tikzpicture}[scale=0.7]
\draw (0,-0) -- (2,-0); 
\draw (0,-1.5) -- (2,-1.5); 
\draw (0,-0) -- (0,-2.25); 
\draw (2,-0) -- (2,-2.25); 
\draw (0,-0) -- (-1.2,-0);
\draw (0,-0) -- (-0,+1.2);
\draw (2,-0) -- (+2.5,+0.5);
\node[] at (-0.7,+0.7) {\rotatebox{45}{$\cdots$}};
\node[dw] at (0,-1.5) {};
\node[db] at (2,-1.5) {};
\draw [thick,double] (0,0) -- (-1,-1);
\node[circle, black, fill=grayn, minimum width=5.6*\onshellradius, draw, inner sep=0pt] at (0,0) {$\scriptstyle \ff_{n-1,3}$};
\node[dw] at (2,0) {};
\node[] at (0,-2.25-\labelvdist) {\footnotesize$1$};
\node[] at (2,-2.25-\labelvdist) {\footnotesize$n$};
\end{tikzpicture}
\end{aligned} \eqncom
\end{equation}
where $n_r=n-n_l+2$. Without loss of generality
we choose to use the common BCFW shift at legs $n$ and $1$.

Recall that bipartite on-shell diagrams are associated with
a decorated permutation $\sigma(i)\geq i$, which can be read off the diagram using left-right paths \cite{ArkaniHamed:2012nw}. The permutation $i\rightarrow \sigma(i)$ is obtained starting from the external leg labelled $i$ and then turning right/left when encountering a black/white vertex (three-point MHV/$\overline{\rm MHV}$ amplitude), ending finally on the external leg $\sigma(i)$. For the purpose of understanding the contours of the Gra\ss mannian integral, we use the fact that this permutation encodes linear relations among the columns $C_i$ in the Gra\ss mannian $G(k,n+2)$, when viewed as $k$-dimensional vectors. These linear relations are sufficient to determine the configuration of points in the Gra\ss mannian $G(k,n+2)$ associated with any on-shell diagram, thus fixing the contour of integration for the associated Graßmannian integral. 

In particular, $\sigma(i)= i+1$ leads to a linear relation between vectors $C_i$ and $C_{i+1}$, while $\sigma(i)= i+2$ gives a linear relation among $C_i, C_{i+1}$ and $C_{i+2}$, rendering these points collinear in projective space. 
For NMHV amplitudes or form factors we consider in this section, the vectors $C_i$ are three-dimensional.
In this case, writing these linear relations in terms of minors, we obtain the following dictionary from permutations to vanishing minors, for any label $ a $, 
\begin{align}
\begin{split}
&    \sigma(i)= i+1 
\quad\implies\quad
(a \ssep i\ssep i\splus1 )=(i\ssep i\splus1 \ssep a)=0\\
&    \sigma(i)= i+2 
\quad\implies\quad
(i\ssep i\splus1 \ssep i\splus2)=0
\eqndot
\label{eq:cond}
\end{split}
\end{align}
In order to apply this strategy to form factors, we first map the form factor diagram to an amplitude diagram by replacing the minimal form factor with a four-point amplitude, as in \cite{Frassek:2015rka},
\begin{equation}
\label{eq:ffamp}
\begin{aligned} 
\begin{tikzpicture}[scale=0.8]
\draw[thick,double] (1.5,-0+1.5) -- (1.5,-0.5+1.5); 
\draw (1.5,-0.5+1.5) -- (2,-\vacuumheight+1.4);  
\draw (1.5,-0.5+1.5) -- (1,-\vacuumheight+1.4);
\end{tikzpicture}
\end{aligned}
\quad
\longleftrightarrow
\quad
\begin{aligned}
\begin{tikzpicture}[scale=0.8]
\draw (1,1) -- (1,2) -- (2,2) -- (2,1) -- (1,1);
\draw (0.5,0.5) -- (1,1);
\draw (2.5,0.5) -- (2,1);
\draw (2.5,2.5) -- (2,2);
\draw (0.5,2.5) -- (1,2);
\node[dw] at (2,1) {};
\node[dw] at (1,2) {};
\node[db] at (1,1) {};
\node[db] at (2,2) {};
\end{tikzpicture}
\end{aligned}
\eqndot
\end{equation} 
This replacement works for reading off the configuration in the Graßmannian because any constraint which does not involve the two columns corresponding to the operator insertion is also present in the purely on-shell part of the form factor diagram, with the minimal form factor removed. This latter diagram, however, has two degrees of freedom fewer, which are restored by the auxiliary four-point amplitude.
\begin{figure}[t]
	\centering
	\begin{tabular}{b{0.33\linewidth}b{0.33\linewidth}b{0.33\linewidth}}
		\small (a) & (b) & (c) \\
		\includegraphics[scale=0.28]{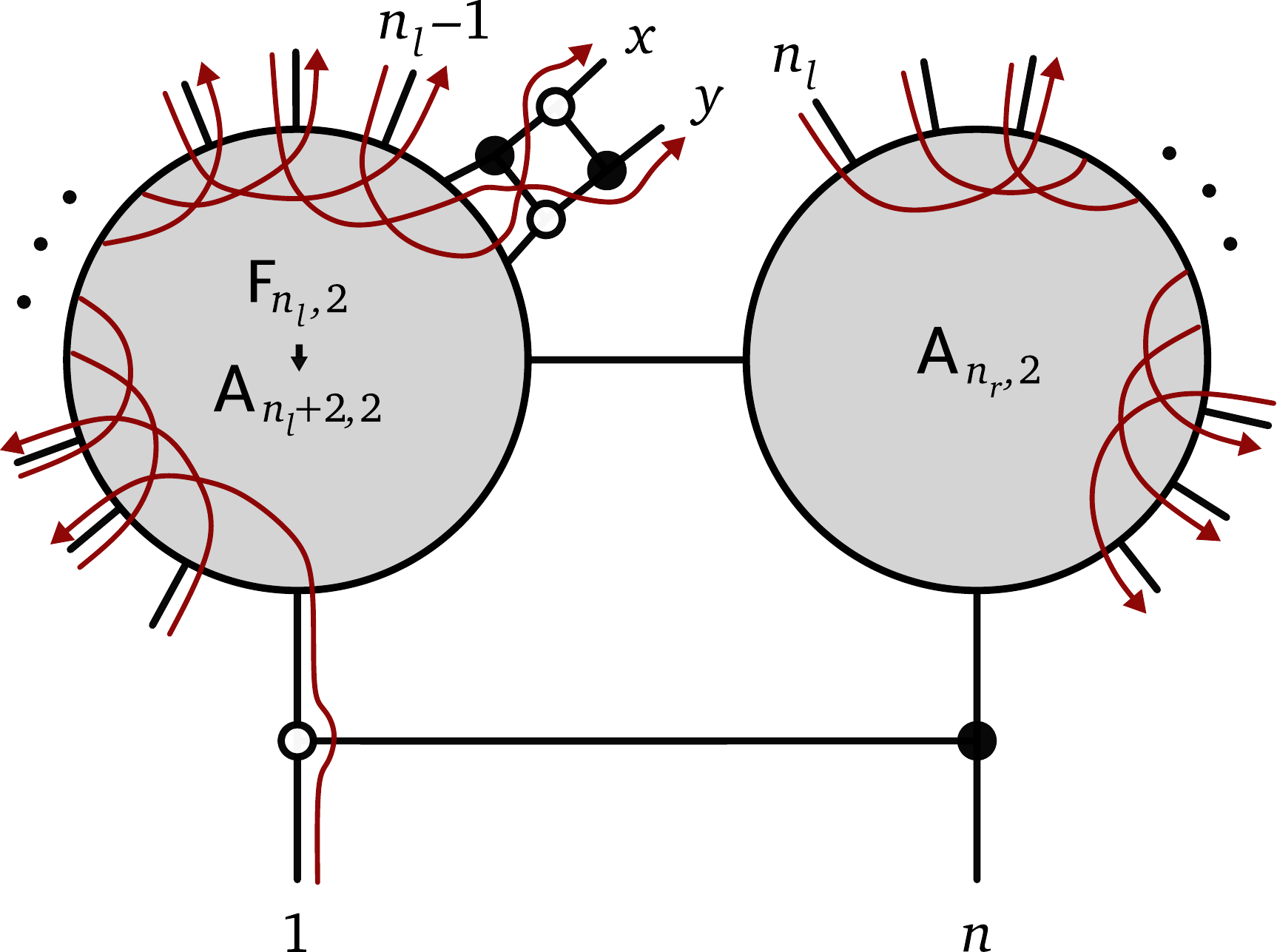}&
		\includegraphics[scale=0.28]{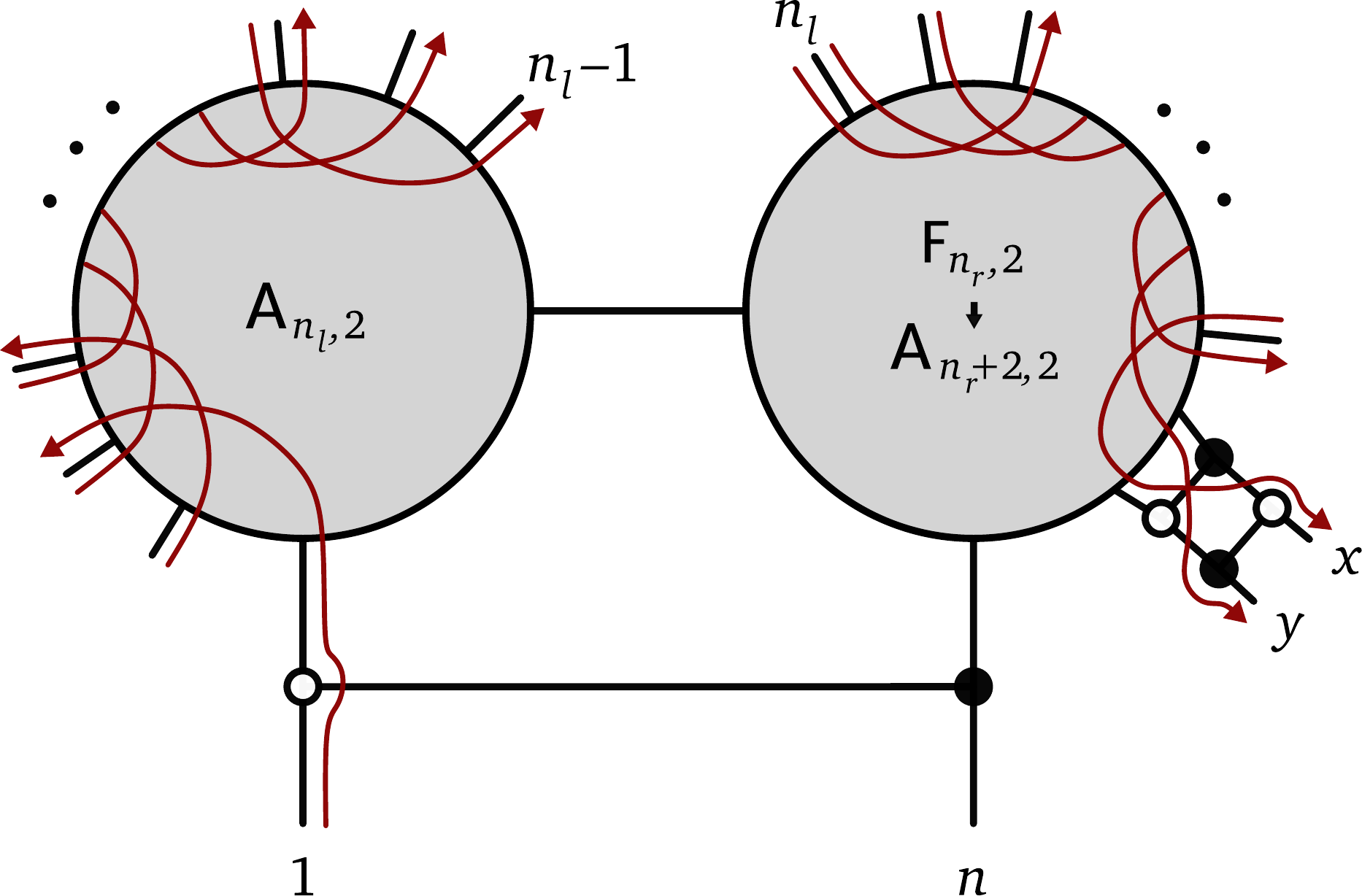}&
		\includegraphics[scale=0.28]{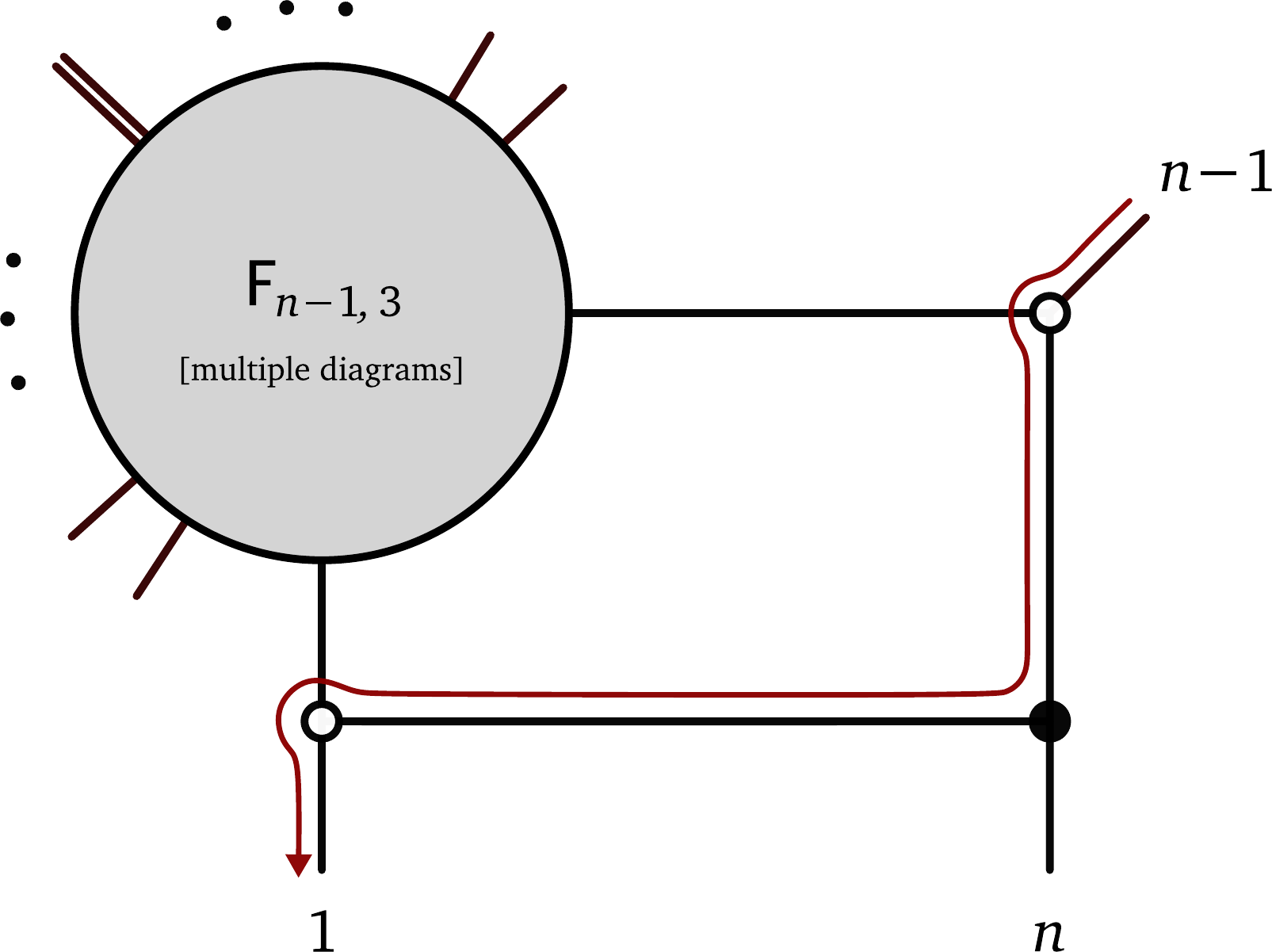}
	\end{tabular}
	\caption{Structure of left-right paths for the three types of terms contributing 
		to the NMHV form factor in the BCFW recursion  relations.}
	\label{fig:terms}
\end{figure}\\[10pt]
We now consider the three types of terms in the BCFW recursion relation \eqref{eq:bcfwnmhv}, depicted in Figure \ref{fig:terms}, in turn.

\paragraph{MHV form factor $\times$ MHV amplitude}
We first work out the configurations for BCFW terms 
with an MHV form factor on the left side of the factorization 
and an MHV amplitude on the right. 
These on-shell diagrams have the form shown in Figure~\ref{fig:terms}(a).
Note that the additional four-point amplitude with labels $x$ and $y$ could have been added between any two of the external labels of the sub-form factor on the left of the diagram because the operator is a colour singlet,
and the sub-form factor therefore cyclically invariant. 
After the transformation $\ff_{n,2}\rightarrow \amp_{n+2,2}$, the two amplitudes in the diagram are MHV and thus the permutations associated with the sub-diagrams are
given by $\sigma_{l/r}(i)=i+2$. Each sub-diagram therefore imposes a geometrical configuration for which the $C_i$ related to its external states all lie on the same line.
More concretely, the sub-form factor ensures 
that $C_1$ up to $C_{n_l-1}$ all lie on the line in $\mathbb{CP}^2$ defined by $x$ and $y$, which we denote by $(xy)$. This results in the vanishing of the minors $\poleu{1}$ up to $\poleu{n_l-1}$:
\begin{equation}
\begin{aligned}
\begin{tikzpicture}
\draw (0,0) -- (4,0);
\node[dot] at (0.5,0) {};
\node[] at (0.5,0.34) {$x$};
\node[dot] at (1.0,0) {};
\node[] at (1.0,0.3) {$y$};
\node[dot] at (1.5,0) {};
\node[] at (1.5,0.345) {$1$};
\node[] at (2.2,0.3) {$\cdots$};
\node[dot] at (3.3,0) {};
\node[] at (3.3,0.32) {$n_l-1$};
\end{tikzpicture}
\end{aligned}
\qquad \longrightarrow \qquad
\poleu{1},\ldots,\poleu{n_l-1}=0\eqndot
\end{equation}
From the MHV amplitude, we can read off the collinearity of $C_{n_l}$ through $C_{n-1}$
which implies that the following minors vanish:
\begin{equation}
\begin{aligned}
\begin{tikzpicture}
\draw (0,0) -- (5,0);
\node[dot] at (0.5,0) {};
\node[] at (0.5,0.3) {$n_l$};
\node[dot] at (1.5,0) {};
\node[] at (1.5,0.33) {$n_l+1$};
\node[] at (2.5,0.33) {$\cdots$};
\node[dot] at (3.3,0) {};
\node[] at (3.3,0.39) {$n-2$};
\node[dot] at (4.3,0) {};
\node[] at (4.3,0.395) {$n-1$};
\end{tikzpicture}
\end{aligned}
\qquad \longrightarrow \qquad
\pole{n_l},\ldots,\pole{n-3} = 0\eqndot
\end{equation}
This gives us $n-3$ residues of the form
\begin{equation}
\{\poleu{1},\ldots,\poleu{n_l-1},
\pole{n_l},\ldots,\pole{n-3}\}
\; ,\quad
\text{for $n_l=2,\ldots,n-2$ }
\eqncom
\end{equation}
which are all the terms with $m=n-3$ in \eqref{eq:FFcontour}, namely  $R_{n-3}^{00}+\sum\nolimits_{i_1=1}^{n-3}  R^{i_10}_{n-3}$.\\[5pt]
As noted above, in order to fully specify a ``contour'', we need to prescribe
which of the cyclically related top forms to use.
Since the MHV sub-form factor is cyclically invariant in its on-shell legs, 
for each term we can take the residue from any top form with a shift of
\begin{equation}
s=0,1,\ldots,n_l-1
\eqndot
\end{equation}
Note that a shift of $s=0$  appears to be incompatible
with our choice of BCFW shift, as the BCFW bridge does not allow the minimal form
factor to be between legs $n$ and $1$. The validity of this shift nevertheless
follows from the consistency of all possible adjacent BCFW shifts.
Moreover, we note that the top forms
with these shifts are exactly those which contain a pole of the given form.

\paragraph{MHV amplitude $\times$ MHV form factor}
The second type of term has the schematic form given in Figure~\ref{fig:terms}(b),
and the argument is similar to the terms just discussed. In particular, the four-point amplitude with $x$ and $y$ could have been attached in other positions for the sub-form factor on the right-hand-side.
In this case, the sub-amplitude and sub-form factor enforce
\begin{align}
&\begin{aligned}
\begin{tikzpicture}
\draw (0,0) -- (3.6,0);
\node[dot] at (0.5,0) {};
\node[] at (0.5,0.3) {$1$};
\node[dot] at (1.0,0) {};
\node[] at (1.0,0.3) {$2$};
\node[] at (1.8,0.3) {$\cdots$};
\node[dot] at (2.8,0) {};
\node[] at (2.8,0.3) {$n_l-1$};
\end{tikzpicture}
\end{aligned}
\qquad \longrightarrow \qquad
\pole{1},\ldots,\pole{n_l-3} = 0
\qquad\text{(sub amplitude)}
\eqncom
\\
&\begin{aligned}
\begin{tikzpicture}
\draw (0,0) -- (4,0);
\node[dot] at (0.5,0) {};
\node[] at (0.5,0.34) {$x$};
\node[dot] at (1.0,0) {};
\node[] at (1.0,0.3) {$y$};
\node[dot] at (1.5,0) {};
\node[] at (1.5,0.325) {$n_l$};
\node[] at (2.2,0.3) {$\cdots$};
\node[dot] at (3.3,0) {};
\node[] at (3.3,0.34) {$n-1$};
\end{tikzpicture}
\end{aligned}
\qquad \longrightarrow \qquad
\poleu{n_l},\ldots,\poleu{n-1}
\;\qquad\text{(sub form factor)}
\eqndot
\end{align}
This gives $n-2$ terms with poles
\begin{equation}
\label{eq:MHVAxMHVFFterms}
\{\pole{1},\ldots,\pole{n_l-3},
\poleu{n_l},\ldots,\poleu{n-1}\}
\; ,\quad
\text{for $n_l=3,\ldots,n$ }
\eqncom
\end{equation}
which are the terms of the form $\sum\nolimits_{m=0}^{n-3} \sum\nolimits_{i_2=1}^{n-m-3}  R^{0i_2}_m$ in \eqref{eq:FFcontour}.\\[5pt]
The possible shifts for these configurations are
\begin{equation}
s=\begin{cases}
n_l-1,\ldots,n-1 & \,\text{ for }n_l=3,\ldots,n-1 \\
0,n-1 & \,\text{ for }n_l=n 
\end{cases}
\eqncom
\end{equation}
which again follow from the cyclicity of the sub-form factor, 
except for the shift $s=0$, which is nevertheless valid and ensures
that all top forms which contain the respective pole can be used
to obtain the corresponding BCFW term.

\paragraph{Lower point NMHV form factor}
The last term in \eqref{eq:bcfwnmhv} is the most interesting one,
since it contains the lower point NMHV form factor $\ff_{n-1,3}$,
which itself is given in terms of a sum of diagrams.
It is the inverse soft limit of this $n-1$ point 
NMHV form factor, with a $k$-preserving inverse soft factor attached to the diagram
as in  Figure~\ref{fig:terms}(c).
For each term in the sub-form factor $\ff_{n-1,3}$, the
inverse soft factor imposes $\pole{n-1}=0$, in addition to the vanishing 
minors of the lower point form factor:
\begin{equation}
\sum_{\mathrm{subdiagrams}} \{\text{poles of subdiagram}\} \cup \{\pole{n-1}\}
\label{eq:recursivepoles}\eqndot
\end{equation}
These terms are the remaining ones in \eqref{eq:FFcontour}, namely $\sum\nolimits_{m=0}^{n-2}
\sum\nolimits_{i_1=1}^m  R^{i_10}_m $. The poles of the sub-diagram are obtained in exactly the same way, meaning that the explicit knowledge of the BCFW poles\footnote{We remark that we use the term ``BCFW pole'' to denote the pole in the Graßmannian integral the residue of which produces a term in the BCFW recursion relation.} for cases with low $n$ is enough to specify the contour for any number of legs recursively. Note that the possible shifts are simply inherited from the sub-diagram
\paragraph{General structure of the NMHV contour} The recursive structure of the contour \eqref{eq:FFcontour} becomes clear if one arranges the residues on a grid, as those shown in Figure~\ref{fig:contourgrid} for $n=4,5,6$. In those pictures the poles corresponding to MHV form factor $\times$ MHV amplitude factorization channels are arranged in the first row and the poles corresponding to MHV amplitude $\times$ MHV form factor channels lie in the last column. Finally, the poles of the form \eqref{eq:recursivepoles} form a sub-grid which obeys the same pattern, but with one point fewer. Using the labels of \eqref{eq:FFcontour}, the rows are sorted with increasing value of $m$, and each row starts with the terms $R^{i_1 0}_m$ with decreasing values of $i_1$ followed by $R^{0 i_2}_m$ with increasing values of $i_2$. 
We also observe that this contour bears similarity to the tree-level contour 
for scattering amplitudes, reviewed in \eqref{eq:ampContour}.
Lastly, note that despite appearing as poles in the Graßmannian integral \eqref{eq:gi}, the general formula for the contour \eqref{eq:FFcontour} never produces a residue at configurations involving an intersection of lines.
\newpage
\begin{figure}[H]
	\newcolumntype{Y}{>{\centering\arraybackslash}X}
	\arrayrulecolor{Blue}
	\vspace*{8pt}
	
	\begin{tabularx}{0.35\textwidth}{YY}
		\multicolumn{2}{c}{$\ff_{4,3}$}\\\arrayrulecolor{black}\hline\arrayrulecolor{Blue}\\
		&
		\\[-16pt]
		$\{\poleu{1}\}$ &
		$\{\pole{1}\}$
		\\
		$s=0,1$ &
		$s=0,3$
		\\
		$\color{Blue}\ff_{2,2}\times\amp_{4,2}$ &
		$\color{Blue}\amp_{4,2}\times\ff_{2,2}$
		\\
		&
		\\
		& 
		\\
		$\{\pole{3}\}$ 
		&
		$\{\poleu{3}\}$ 
		\\
		$s=1,2$
		& 
		$s=2,3$
		\\
		$\color{Blue}\ff_{3,3}\times\amp_{3,1}$
		& 
		$\color{Blue}\amp_{3,2}\times\ff_{3,2}$
	\end{tabularx}
	\hfill
	\begin{tabularx}{0.5\textwidth}{YYY}
		\multicolumn{3}{c}{$\ff_{5,3}$}\\\arrayrulecolor{black}\hline\arrayrulecolor{Blue}\\
		&&
		\multicolumn{1}{:c}{}
		\\[-16pt]
		$\{\poleu{1},\poleu{2}\}$ &
		$\{\poleu{1},\pole{2}\}$ &
		\multicolumn{1}{:c}{
			$\{\pole{1},\pole{2}\}$
		} \\
		$s=0,1,2$ &
		$s=0,1$ &
		\multicolumn{1}{:c}{
			$s=0,4$
		} \\
		$\color{Blue}\ff_{3,2}\times\amp_{4,2}$ &
		$\color{Blue}\ff_{2,2}\times\amp_{5,2}$ &
		\multicolumn{1}{:c}{
			$\color{Blue}\amp_{5,2}\times\ff_{2,2}$
		} \\
		&&
		\multicolumn{1}{:c}{
		} \\
		\cline{1-2}
		\multicolumn{1}{|c}{
		} & 
		& \multicolumn{1}{|c}{
		}\\
		\multicolumn{1}{|c}{
			$\{\poleu{1},{\color{Blue}\pole{4}}\}$ 
		} & 
		$\{\pole{1},{\color{Blue}\pole{4}}\}$ 
		& \multicolumn{1}{|c}{
			$\{\pole{1},\poleu{4}\}$ 
		}\\
		\multicolumn{1}{|c}{
			$s=0,1$
		} & 
		$s=0,3$
		& \multicolumn{1}{|c}{
			$s=3,4$
		}\\
		\multicolumn{1}{|c}{
		} & 
		& \multicolumn{1}{|c}{
			$\color{Blue}\amp_{4,2}\times\ff_{3,2}$
		}\\
		\multicolumn{1}{|c}{
		} & 
		& \multicolumn{1}{|c}{
		}\\
		\multicolumn{1}{|c}{
			$\{\pole{3},{\color{Blue}\pole{4}}\}$ 
		} & 
		$\{\poleu{3},{\color{Blue}\pole{4}}\}$ 
		& \multicolumn{1}{|c}{
			$\{\poleu{3},\poleu{4}\}$ 
		}\\
		\multicolumn{1}{|c}{
			$s=1,2$
		} & 
		$s=2,3$
		& \multicolumn{1}{|c}{
			$s=2,3,4$
		}\\
		\multicolumn{1}{|c}{
		} & 
		& \multicolumn{1}{|c}{
			$\color{Blue}\amp_{3,2}\times\ff_{4,2}$
		}\\
		\cline{1-2}
		\multicolumn{1}{l}{
			{\color{Blue}$\ff_{4,3}\times\amp_{3,1}$}
		} & 
		& \multicolumn{1}{c}{
		}\\
	\end{tabularx}
	
	\vspace{3.5\baselineskip}\hspace{1.6cm}
	\begin{tabularx}{0.8\textwidth}{YYYY}
		\multicolumn{4}{c}{$\ff_{6,3}$}\\\arrayrulecolor{black}\hline\arrayrulecolor{Blue}\\
		&&& 
		\multicolumn{1}{:c}{}
		\\[-16pt]
		$\{\poleu{1},\poleu{2},\poleu{3}\}$ &
		$\{\poleu{1},\poleu{2},\pole{3}\}$ &
		$\{\poleu{1},\pole{2},\pole{3}\}$ &
		\multicolumn{1}{:c}{
			$\{\pole{1},\pole{2},\pole{3}\}$
		} \\
		$s=0,1,2,3$ &
		$s=0,1,2$ &
		$s=0,1$ &
		\multicolumn{1}{:c}{
			$s=0,5$
		} \\
		$\color{Blue}\ff_{4,2}\times\amp_{4,2}$ &
		$\color{Blue}\ff_{3,2}\times\amp_{5,2}$ &
		$\color{Blue}\ff_{2,2}\times\amp_{6,2}$ &
		\multicolumn{1}{:c}{
			$\color{Blue}\amp_{6,2}\times\ff_{2,2}$
		} \\
		&&&
		\multicolumn{1}{:c}{
		} \\
		\cline{1-3}
		\multicolumn{1}{|c}{
		} & 
		&
		& \multicolumn{1}{|c}{
		}\\
		\multicolumn{1}{|c}{
			$\{\poleu{1},\poleu{2},{\color{Blue}\pole{5}}\}$ 
		} & 
		$\{\poleu{1},\pole{2},{\color{Blue}\pole{5}}\}$ 
		&
		$\{\pole{1},\pole{2},{\color{Blue}\pole{5}}\}$ 
		& \multicolumn{1}{|c}{
			$\{\pole{1},\pole{2},\poleu{5}\}$ 
		}\\
		\multicolumn{1}{|c}{
			$s=0,1,2$
		} & 
		$s=0,1$
		&
		$s=0,4$
		& \multicolumn{1}{|c}{
			$s=4,5$
		}\\
		\multicolumn{1}{|c}{
		} & 
		&
		& \multicolumn{1}{|c}{
			$\color{Blue}\amp_{5,2}\times\ff_{3,2}$
		}\\
		\multicolumn{1}{|c}{
		} & 
		&
		& \multicolumn{1}{|c}{
		}\\
		\multicolumn{1}{|c}{
			$\{\poleu{1},\pole{4},{\color{Blue}\pole{5}}\}$ 
		} & 
		$\{\pole{1},\pole{4},{\color{Blue}\pole{5}}\}$ 
		&
		$\{\pole{1},\poleu{4},{\color{Blue}\pole{5}}\}$ 
		& \multicolumn{1}{|c}{
			$\{\pole{1},\poleu{4},\poleu{5}\}$ 
		}\\
		\multicolumn{1}{|c}{
			$s=0,1$
		} & 
		$s=0,3$
		&
		$s=3,4$
		& \multicolumn{1}{|c}{
			$s=3,4,5$
		}\\
		\multicolumn{1}{|c}{
		} & 
		&
		& \multicolumn{1}{|c}{
			$\color{Blue}\amp_{4,2}\times\ff_{4,2}$
		}\\
		\multicolumn{1}{|c}{
		} & 
		&
		& \multicolumn{1}{|c}{
		}\\
		\multicolumn{1}{|c}{
			$\{\pole{3},\pole{4},{\color{Blue}\pole{5}}\}$ 
		} & 
		$\{\poleu{3},\pole{4},{\color{Blue}\pole{5}}\}$ 
		&
		$\{\poleu{3},\poleu{4},{\color{Blue}\pole{5}}\}$ 
		& \multicolumn{1}{|c}{
			$\{\poleu{3},\poleu{4},\poleu{5}\}$ 
		}\\
		\multicolumn{1}{|c}{
			$s=1,2$
		} & 
		$s=2,3$
		&
		$s=2,3,4$
		& \multicolumn{1}{|c}{
			$s=2,3,4,5$
		}\\
		\multicolumn{1}{|c}{
		} & 
		&
		& \multicolumn{1}{|c}{
			$\color{Blue}\amp_{3,2}\times\ff_{5,2}$
		}\\
		\cline{1-3}
		\multicolumn{1}{l}{
			{\color{Blue}$\ff_{5,3}\times\amp_{3,1}$}
		} & 
		&
		& \multicolumn{1}{c}{
		}\\
	\end{tabularx}
	
	\vspace{3\baselineskip}

	\caption{
		Poles contributing to the four, five and six-point NMHV form factors.
		The corresponding factorization channels are indicated in blue,
		and we list all possible values of the shift $s$, which label the
		Graßmannian top forms featuring the respective pole.
		The blue boxes contain the poles from the inverse soft limit of the
		lower-point form factor, which share the vanishing minor $\pole{n-1}$ appended to the contour of $\ff_{n-1,3}$.
	}
	\label{fig:contourgrid}
\end{figure}
\newpage
We also note that, although \eqref{eq:FFcontour} cannot generally be thought of as a contour in the real sense, a special case where \eqref{eq:FFcontour} \emph{can} be interpreted as such is for $n=4$. Summing two top forms \eqref{eq:gi} with shifts $s=0$ and $s=2$ subject to the contour  \eqref{eq:FFcontour} we get
\begin{multline}
\Big[\gi_{4,3}^{[0]} + \gi_{4,3}^{[2]}\Big]\bigg|_{\contour_{4,3}} = 
\abr{xy}^2
\int_{\contour_{4,3}}\frac{\dd^{3\times 6}C}{\text{Vol}[GL(3)]}\;
\delta^{2\times 3}(C\cdot\vlt) \, \delta^{4\times 3}(C\cdot\vle) \, \delta^{2\times(n-1)}(C^\perp\cdot\vll)
\\
\times \left[\frac{1}{\big[
	\pole{1}\pole{2}\poleu{1}\poleu{4}  (xy\,(34)\cap(12))
	\big]}+\frac{1}{\big[
	\pole{3}\pole{4}\poleu{3}\poleu{2}  (xy\,(12)\cap(34))
	\big]}\right]\ .
\label{eq:gi4}
\end{multline}
\noindent
According to \eqref{eq:FFcontour}, the contour for $n=4$ is
\begin{equation}
\contour_{4,3} = \{\pole{1}\}+ \{\pole{3}\}+\{\poleu{1}\}+\{\poleu{3}\}\, =\, -\big[\{\pole{2} \} + \{\pole{4} \}+\{\poleu{2} \} + \{\poleu{4} \} + \{\pole{xy\,(12)\cap(34)} \}\big]\eqncom 
\end{equation}
where in the last line we have used Cauchy's theorem. Interestingly, the combination $\{\pole{2} \} + \{\pole{4} \}+\{\poleu{2} \} + \{\poleu{4} \}$ gives the (P)BCFW contour and for the residue $\{\pole{xy\,(12)\cap(34)} \}$ the contributions of the two top forms cancel out.
The fact that the integrands can be combined in this way is accidental for $n=4$ since the top forms with $s=0$ and $s=2$ together contain all poles contributing to the BCFW representation. Therefore \eqref{eq:gi4} returns the form factor. For larger values of $n$, as can be seen by inspecting Figure \ref{fig:contourgrid}, there is no combination of top forms which contains all poles picked out by the contour the same number of times, and therefore a combination such as \eqref{eq:gi4} is not possible. 

It is clear that the prescription given above for obtaining contours applies to general N$^{k-2}$MHV form factors. Just like for scattering amplitudes \cite{ArkaniHamed:2012nw}, the contour of a given form factor is determined by the on-shell diagrams dictated by the BCFW recursion relation. We showed explicitly for NMHV form factors the general property that the decorated permutations of the corresponding bipartite on-shell diagrams provide the necessary information to select the lower-dimensional cells
of the Graßmannian $G(k,n+2)$ which contribute to a general $n$-point N$^{k-2}$MHV form factor.

For scattering amplitudes, there exists a second way of obtaining the general tree-level contour for the Graßmannian integral in a compact closed form, namely the connected prescription \cite{Bourjaily:2010kw} derived from the twistor string. In the following sections, we study the analogous connected formula for form factors \cite{He:2016jdg, Brandhuber:2016xue}.

\section{A Graßmannian formulation from the connected prescription}
\label{sec:connected}
So far we have considered the $G(3,n+2)$ formulation of form factors which is analogous to the $G(3,n)$ amplitudes formula \eqref{eq:ampG}, namely a contour integral equipped with a tree-level contour \cite{ArkaniHamed:2009dn}.
A dual $G(3,n)$ formulation for scattering amplitudes arises from the connected formula after the embedding of $G(2,n)$ into $G(3,n)$ \cite{Nandan:2009cc,ArkaniHamed:2009dg}. This mapping returns a representation of the $G(3,n)$ integral which by construction inherits the contour of the connected formula. 
This section is devoted to studying the analogous connected formula for form factors. In particular, we present a lift from the link representation of \cite{Brandhuber:2016xue} to the Graßmannian valid for any value of $n$.

\subsection{Brief review of the connected prescription and link representation}

\label{sec:connected-to-link}

In analogy with the amplitude connected prescription \cite{Roiban:2004yf}, in \cite{Brandhuber:2016xue} and \cite{He:2016jdg} a similar formula was obtained for form factors of the chiral part of the stress tensor operator. 
This representation was given an ambitwistor string interpretation in \cite{Brandhuber:2016xue} and \cite{Bork:2017qyh}. Here we review the derivation of \cite{Brandhuber:2016xue} for the form factor connected formula in the link representation. The kinematic setup is the same as for the Graßmannian integral: we add to the set of $n$ on-shell states two additional particles labelled by $x$ and $y$, representing the kinematics of the operator. Then, for a helicity sector with Graßmann degree $4k$ one chooses $k$ labels from the set $\{1,\dots,n\}$ to form the set ${\rm m}$, indexed by upper case letters $I=\{i_1,\dots, i_k\}$. The remaining $n+2-k$ labels (which always contain $x$ and $y$) form the set $\overline{\rm p}$, labelled by lower case letters $i$. The set $\rm p$ is the same as $\overline{\rm p}$ with $x$ and $y$ removed.\\[10pt]
Using this notation, the form factor connected formula reads 
\begin{align}
\label{eq:ff-connected}
\begin{split}
\ff_{n,k} = &\b{xy}^2 \int \frac{1}{\text{Vol}(GL(2))} \frac{\dd^2\sigma_x \dd^2\sigma_y}{(xy)^2}\prod_{a=1}^n \frac{\dd^2\sigma_a}{(a\,a+1)}\\
&\times \prod_{i\in \overline{\rm p}} \delta^2 (\lambda_i - \lambda(\sigma_i)) \prod_{I \in {\rm m}} \delta^{2} (\vlt_I - \vlt(\sigma_I)) \delta^{4} (\vle_I - \vle(\sigma_I))\ ,
\end{split}
\end{align}
where 
$(\sigma_a^1, \sigma_a^2)$
are homogeneous coordinates in $\mathbb{CP}^1$,  $(ab)=\epsilon_{\alpha \beta} \sigma_a^\alpha \sigma_b^\beta$, and  
\begin{align}
\label{eq:Witten-RSV}
\lambda(\sigma_I) = \sum_{i \in \overline{\rm p}} \frac{1}{(Ii)}\lambda^i,\quad \vlt(\sigma_i)=-\sum_{I \in {\rm m}} \frac{1}{(Ii)}\vlt^I,\quad \vle(\sigma_i)=-\sum_{I \in {\rm m}} \frac{1}{(Ii)}\vle^I\ .
\end{align}

As is the case with scattering amplitudes, one can go from the connected prescription to the \emph{link representation} by introducing a new set of variables $c_{Ij}$, termed \emph{link variables} \cite{ArkaniHamed:2009si}, and imposing the additional equations $c_{Ij} = \frac{1}{(Ij)}$  \cite{Spradlin:2009qr,Dolan:2009wf}.
The advantage of using these variables is that the equations \eqref{eq:Witten-RSV} become linear.
In \cite{Brandhuber:2016xue}, a generic expression for form factors in this representation was given:
\begin{align}
\label{eq:ff-link}
\ff_{n,k}= &\b{xy}^2 \int \prod_{I \in {\rm m}, j \in  \overline{\rm p}} \dd c_{Ij} U(c_{Ij}) \times \prod_{i\in \overline{\rm p}} \delta^2 (\lambda_i - c_{Ii}\lambda_i) \prod_{I \in {\rm m}} \delta^{2} (\vlt_I + c_{Ii} \vlt_i) \delta^{4} (\vle_I + c_{Ii} \vle_i) \\
\label{eq:U}
&U(c_{Ii})= \int \frac{1}{\text{Vol}(GL(2))} \frac{\dd^2\sigma_x \dd^2\sigma_y}{(xy)^2} \prod_{a=1}^n \frac{\dd^2\sigma_a}{(a\,a+1)} \prod_{I \in {\rm m}, i \in  \overline{\rm p}} \delta\left(c_{Ii}-\frac{1}{(Ii)}\right) \ .
\end{align}
Note that although \eqref{eq:ff-link} carries the degrees of freedom of a $G(k,n+2)$ Graßmannian formula, all integration variables are fixed by the delta functions.
Similarly to what was done for scattering amplitudes in \cite{ArkaniHamed:2009dn}, we now lift this formulation in the NMHV case to a fully $GL(3)$ invariant Graßmannian formulation by performing the $\sigma$ integrations.

\subsection{From the link representation to the Graßmannian}

In the following we focus on our case of interest, namely NMHV form factors with $k=3$,
and write \eqref{eq:ff-link} with the integrand \eqref{eq:U} in the form of a $GL(3)$ invariant Graßmannian integral, with no free 
integration variables.
Indeed, while the explicit delta functions of \eqref{eq:ff-link} can only fix  $2n$  out of the $3(n-1)$ integration variables $c_{Ii}$, the function $U(c_{Ij})$ provides precisely the additional $n-3$ constrains required to solve for all $c_{Ij}$.
After solving $2n$ out of the $3n-3$ constraints imposed by the delta functions of \eqref{eq:U},
there are no integrations over the variables $\sigma_a$ left.
It is then straightforward to restore the $GL(3)$ invariance.

The $n-3$ remaining delta functions, evaluated at the solutions of the others, generate constraints depending on six points each. These equations, when written in terms of $GL(3)$ minors, have the general form $\delta(S_{i_1 i_2 i_3 i_4 i_5 i_6})$, where\footnote{This expression is invariant under permutations of the six labels up to a sign of the signature of the permutation.}
\begin{align}
\label{eq:sextic}
S_{i_1 i_2 i_3 i_4 i_5 i_6} \equiv (i_1 i_2 i_3 )(i_3i_4i_5)(i_5i_6 i_1)( i_2 i_4 i_6)-( i_2 i_3 i_4)( i_4 i_5 i_6)( i_6 i_1 i_2)( i_3 i_5 i_1)\ .
\end{align}
The equations $S=0$ are the same that feature for scattering amplitudes, and are in general polynomials of degree four in the link variables. Their geometric meaning was discussed in \cite{White:1915,ArkaniHamed:2009dg}; the localization of N$^{k-2}$MHV scattering amplitudes on degree $(k-1)$-curves in twistor space, as in Witten's twistor string theory, has a counterpart as a localization in the Graßmannian. Namely, by viewing each column in the matrix $C \in G(k,n+2)$  as a point in $\mathbb{CP}^{k-1}$, each column must be the image of a map $\mathbb{CP}^1\mapsto\mathbb{CP}^{k-1}$, generally given by the Veronese map
\begin{align}
\label{eq:veronese}
(\sigma^1, \sigma^2)
\mapsto 
\left(
(\sigma^1)^{k-1} , (\sigma^1)^{k-2}\sigma^2 , \cdots , \sigma^1(\sigma^2)^{k-2} , (\sigma^2)^{k-1}
\right)
\ .
\end{align}
For $k=3$ this corresponds to a map of degree two, and therefore the constraints arising from \eqref{eq:U} must ensure that all $n+2$ points lie on the same curve. This is achieved by a combination of equations of the form \eqref{eq:sextic}, which impose that a sixth point lies on the degree-two curve generated by the other five. For this reason, we refer to these equations as \emph{conic constraints}. It is straightforward to see that, if a matrix $C\in G(3,n+2)$ has all columns as in \eqref{eq:veronese}, all equations \eqref{eq:sextic} trivially vanish since the $3\times 3$ minors factorise in terms of $2\times 2$ minors formed of the $\sigma$ coordinates as $(abc)=(ab)(bc)(ca)$.

Performing an explicit lift of \eqref{eq:ff-link} from the link representation to an integral over $GL(3)$ for low values of $n$ reveals a recursive structure in which the $n$-point form factor is obtained from the $(n-1)$-point as follows:
\begin{equation}
\begin{aligned}
\ff_{n,3} &= \abr{xy}^2 \int \frac{\dd^{3\times (n+2)}C}{\text{Vol(GL(3))}}
\; I_{n,3} \;
\delta^{2\times 3}(C\cdot\vlt) \, \delta^{4\times 3}(C\cdot\vle) \, \delta^{2\times(n-1)}(C^\perp\cdot\vll) 
\  ,\\[10pt]
I_{4,3} &= \frac{(13x)(13y)}{(123)(134)(1xy)(3xy)} \delta(S_{1234xy})\ ,\\[10pt]
I_{n,3} &= I_{n-1,3} \times \left[(-1)^{n-1} \frac{(12n-1)(13n-1)(1xy)(23x)(23y)}{(1n-1n)(23n-1)} \delta(S_{123nxy})\right],\quad n\geq 5\eqncom
\end{aligned}
\label{eq:Fn}
\end{equation}
where we chose to display only the integrands with $n\geq 4$, which are genuinely NMHV.
Although the integrands of this formulation no longer enjoy the manifest cyclic invariance of the connected formula, the conic constraints imposed by the delta functions ensure this symmetry is present.
\\[10pt]
There are several ways of representing the integrand of \eqref{eq:Fn}, all coinciding on the support of the conic constraints. Likewise, the choice of equations appearing inside the delta functions is not unique as the geometric constraint that the $n+2$ points lie on the same degree-two curve can be represented is various distinct ways.
For the particular representation in \eqref{eq:Fn}, we consider the conic defined by the five points $\{1,2,3,x,y\}$ and each conic constraint imposes that one of the other points $\{4,\dots,n\}$ lie on the same curve, as can be seen from the additional constraints present in each recursive factor. The minors appearing in the numerator of the recursive factor are responsible for annihilating spurious solutions of the conic constraints. 
For instance, a configuration where four out of the points belonging to the set $\{1,2,3,x,y\}$ are collinear would set to zero all conic constraints, but would not imply that all points lie on the same curve. The numerator factor $(13x)(13y)(23x)(23y)(1xy)$ precisely vanishes for every configuration of this sort. A special case where the cancellation of spurious solutions of the conic constrains does \emph{not} happen is for $n=5$, since the factor of $(1xy)$ cancels between $I_{4,3}$ and the recursive factor in \eqref{eq:Fn}. In this case, one needs to ensure that only the physical solutions of the conic constraints are taken into account.
This situation is discussed in further detail in Section~\ref{sec:5points}.

\subsection{Formulation with inverse soft interpretation}

For scattering amplitudes, it is possible to interpret the recursive factors $I_n/I_{n-1}$ as the addition of a particle via an inverse soft factor \cite{ArkaniHamed:2009dg,Bourjaily:2010kw}.
The same should be true for form factors, as they are inverse soft 
constructible \cite{Nandan:2012rk}. In particular, one can show that for form factors with sufficiently many on-shell legs, namely six, the effect of the operator may be omitted and it is possible to write the recursive factor of \eqref{eq:Fn} in the same way as that for amplitudes.
This is achieved by rewriting \eqref{eq:Fn} in a way more similar to the amplitude formulas presented in e.g. \cite{Bourjaily:2010kw} by means of the identity
\begin{equation}
\delta(S_{ijkrst})\delta(S_{ijkrsu})
=
\frac{(jkt)(irt)}{(jks)(irs)}
\delta(S_{ijkrst})\delta(S_{ijkrtu})
\eqndot
\label{eq:id}
\end{equation}
We start by considering the ratio $I_{5,3}/I_{4,3}$, and trade $S_{123xy5}\rightarrow S_{123x45}$
on the support of $S_{1234xy}=0$ using \eqref{eq:id}, which results in
\begin{equation}
I_{5,3}/I_{4,3}
=
\frac{(124)(134)(23x)(1x4)}{(145)}
\delta(S_{123x45})
\eqndot
\label{eq:isf5}
\end{equation}
This factor is already much more similar to the amplitude ``soft factor'', but it is clear that either $x$ or $y$, representing the kinematics of the operator, has to be an index in the left-over $S$.
Next we consider $I_{6,3}/I_{5,3}$. We first trade $y$ in $S_{123xy6}$ for $4$ using $S_{1234xy}$, and then $x\rightarrow 5$ using $S_{123x45}$, getting
\begin{equation}
I_{6,3}/I_{5,3}\sim
\frac{(125)(135)(234)(145)}{(156)}
\delta(S_{123456})\eqncom
\end{equation}
which is precisely the recursive factor which maps $\amp_{5,3}$ to $\amp_{6,3}$.

We can now proceed recursively, and find that also for higher point
form factors the recursive structure of the integrand can be written
in exactly the same way as for amplitudes,
\begin{equation}
I_{n,3}/I_{n-1,3} = 
\frac{(12n-1)(13n-1)(1n-2n-1)(23n-2)}{(1n-1n)}
\delta(S_{123n-2n-1n})
\eqncom
\qquad n\geq 6
\eqndot
\label{eq:isfactor}
\end{equation}
This form of the recursive factor is the same as the one used in \cite{ArkaniHamed:2009dg}, 
where it was shown that this factor ensures the correct soft limit for particle $n$.
This representation was also important for matching the connected formula with 
the Graßmannian integral via applications of the GRT, as its integrand has singularities at all BCFW poles.
In the next section we investigate this strategy for form factors. 

\section{From the connected prescription to BCFW via the GRT}
\label{sec:relation}

In the previous sections, we studied two different Graßmannian representations of form factors. On one side there is the formula associated with the BCFW recursion relation and on-shell diagrams, given in \eqref{eq:gi} and equipped with the contour \eqref{eq:FFcontour}. On the other hand there is the formula that arises from the connected prescription, represented as in \eqref{eq:Fn} or \eqref{eq:isfactor}, which does not require a separate specification of the contour.

These formulations are the form factor analogues of corresponding expressions for scattering amplitudes, whose NMHV Graßmannian formulas are related as shown below in Figure \ref{fig:rel} \cite{Nandan:2009cc,ArkaniHamed:2009dg}.
\begin{figure}[h]
	\centering
	\includegraphics[width=0.95\linewidth]{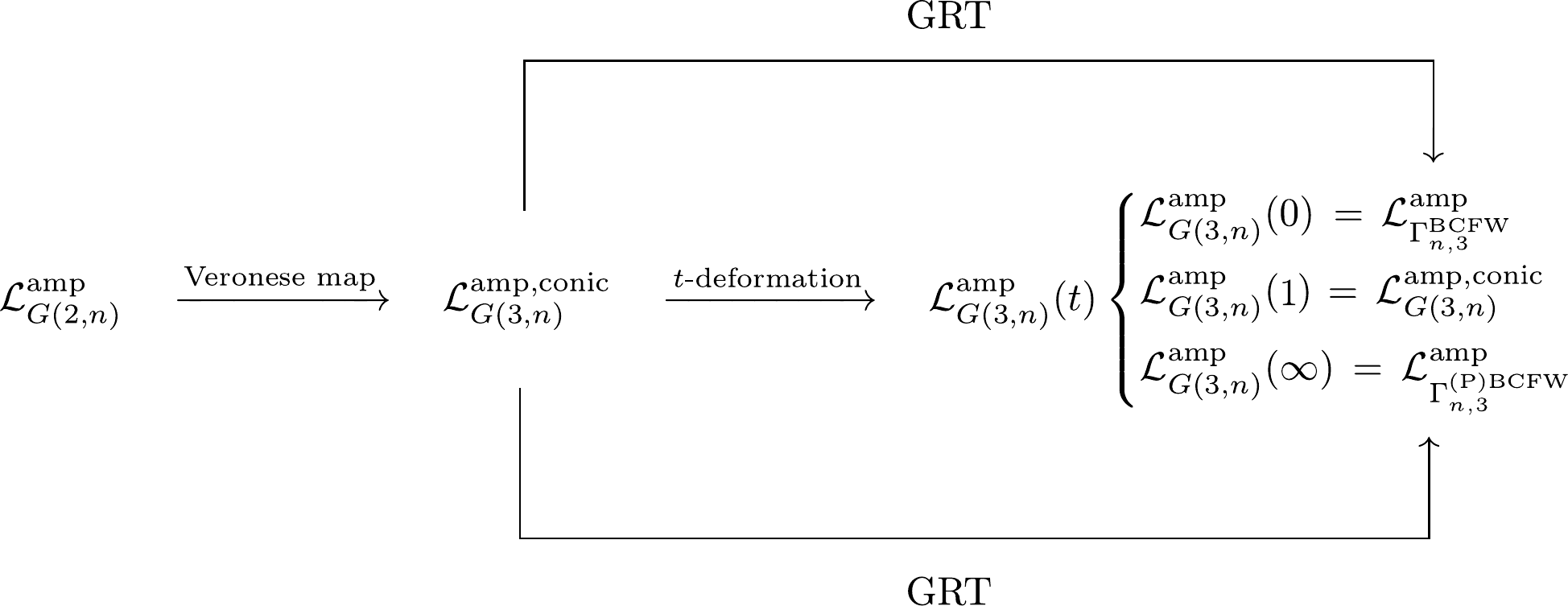}
	\vspace{\baselineskip}
	\caption{
		Relations between different Graßmannian formulations of scattering amplitudes.
		Here $\mathcal{L}_{G(2,n)}^{\mathrm{amp}}$ denotes the amplitude connected formula,
		which can be understood as an integral over the Graßmannian $G(2,n)$.
		The Veronese map leads from $\L^{\rm amp}_{G(2,n)}$ to the Graßmannian integral with conic constraints,  
		$\mathcal{L}_{G(3,n)}^{\mathrm{amp,conic}}$.
		There are different ways in which the Graßmannian integral with BCFW
		or (P)BCFW integration contour,
		$\mathcal{L}_{\Gamma_{n,3}}^{\mathrm{amp}}$, 
		can be obtained from this representation: either via the smooth deformations
		of the conic constraints $\mathcal{L}_{G(3,n)}^{\mathrm{amp,conic}}(t)$,
		or via the application of GRTs.
	}
	\label{fig:rel}
\end{figure}
The Veronese map referred to in this diagram is given in \eqref{eq:veronese}. The $t$-deformation amounts to introducing $n-5$ parameters $t_j$ into the conic constraints \eqref{eq:sextic} in a systematic way, by defining
\begin{align}
\label{eq:sextic-t_j}
S_{i_1 i_2 i_3 i_4 i_5 i_6}(t_j) \equiv (i_1 i_2 i_3 )(i_3i_4i_5)(i_5i_6 i_1)( i_2 i_4 i_6)-t_j\,( i_2 i_3 i_4)( i_4 i_5 i_6)( i_6 i_1 i_2)( i_3 i_5 i_1)\ .
\end{align}
Note in particular that the BCFW contour \eqref{eq:ampContour} can be recovered both from taking limits of the deformation parameters $t_j$ or through applications of the GRT starting from the formula with the conic constraints.

The aim of this section is to investigate the validity of similar relations between the corresponding formulas for form factors.
A preliminary attempt to use the Veronese map to relate the Graßmannian integral based on on-shell diagrams directly to the connected formula was made in \cite{Brandhuber:2016xue}, and found to be impossible. 
Based on the derivation of Section~\ref{sec:contour}, we conclude that the BCFW contour contains poles originating from different top forms in such a way that no linear combination of top forms gives the tree-level form factor with a single contour of integration. Such a single integral, would however be necessary for a direct application of the Veronese map.

In this section,  we explore the possibility of relating the Graßmannian formulations directly using the GRT, focusing on low-point examples. 
Already at four points we find that there is no naive analogue of the $t$-deformation \eqref{eq:sextic-t_j} for the form-factor formulas. Moreover, we show that successive applications of the GRT lead from the Graßmannian formula with conic constraints to that with the BCFW contour for four and five points. However, this is no longer the case starting at six points. We furthermore highlight subtleties involved in the computation of the BCFW residues which do not appear for scattering amplitudes, such as the necessity of regularising residues with a 0/0 behaviour.

\subsection{Four points}

Consider the integral given in \eqref{eq:Fn}, which we repeat here for convenience: 
\begin{equation}
I_{4,3} = \frac{(13x)(13y)}{(123)(134)(1xy)(3xy)} \delta(S_{1234xy})\eqndot
\end{equation}
The contour is defined by the equation $S_{1234xy}=0$. Applying the residue theorem one obtains a new combination of residues given by
\begin{equation}
\label{eq:GRT-4points} 
\{S_{1234xy}\} \rightarrow -\{(123)\}-\{(341)\}-\{(1xy)\}-\{(3xy)\} \eqndot
\end{equation}
The location of these poles are the same as the four-point BCFW contour which can be read off Figure \ref{fig:contourgrid}, cf.~\eqref{eq:poles} for the notation. For each of the factors on the right-hand-side of \eqref{eq:GRT-4points}, the factor of $S_{1234xy}$ in the denominator factorises into a product of four minors.
It is straightforward to check that the value of each residue is the same as that stemming from the Graßmannian formula \eqref{eq:gi}.

A lesson can be taken from this simple case. Consider the analogous example of the six-point scattering amplitude:
\begin{align}
I^{\rm amp}_{6,3} = \frac{(135)}{(123)(345)(561)} \delta(S_{123456}) = \frac{(246)}{(234)(456)(612)} \delta(S_{123456}) . 
\end{align}
In this situation $S_{123456}$ always factorises in the same way for all three poles present in the integrand, both in the BCFW or (P)BCFW representations. This means that one can introduce a parameter $t$ to the term that vanishes as in \eqref{eq:sextic-t_j}, i.e. 
$S_{123456}(t)=
t(123)(345)(561)(246)-(234)(456)(612)(351)$,
and the amplitude is independent of the value of $t$ \cite{ArkaniHamed:2009dg,Nandan:2009cc}. In particular, a one-parameter family of dual Graßmannian theories is defined in this fashion, with the particular cases of the twistor string for $t=1$ and the BCFW and (P)BCFW cases for $t=0$ or $t=\infty$, respectively, as shown schematically in Figure \ref{fig:rel}.

For form factors this is not possible: in the four-point example
we see that $S_{1234xy}$ always factorises, but differently at each pole. Explicitly, using the permutation invariance of the conic constraints in its labels, 
\begin{align}
S_{1234xy} =
\begin{cases}
\phantom{-}S_{314yx2}\quad &\rightarrow\quad\phantom{-}(314)(4yx)(x23)(1y2) \qquad\text{ on }\{(123)\} \\
-S_{312yx4}\quad &\rightarrow\quad -(312)(2yx)(x43)(1y4) \qquad\text{ on }\{(341)\} \\
-S_{243yx1}\quad &\rightarrow\quad-(243)(3yx)(x12)(4y1) \qquad\text{ on }\{(1xy)\} \\
\phantom{-}S_{xy1423} \quad&\rightarrow\quad \phantom{-}(xy1)(142)(23x)(y43) \qquad\text{ on }\{(3xy)\} 
\end{cases} . 
\end{align}
This means that there is no deformation---or at least no naive one---of $S_{1234xy}$ which could interpolate between the Graßmannian integral related to on-shell diagrams and the one based on the connected prescription.\footnote{%
Aspects of this deformation play an important role in the derivation of similar integrals for form factors of Wilson line operators from the ambitwistor string in \cite{Bork:2017qyh}. It would be very interesting to see if the approach of this work can shed more light on this issue.
} 
\subsection{Five points}
\label{sec:5points}

We now consider the five-point form factor, for which the integrand in the inverse soft formulation \eqref{eq:isf5} reads
\begin{equation}
\label{eq:i5}
I_{5,3}=\frac{(13x)(13y)(23x)(124)(14x)}{(123)(1xy)(3xy)(145)} \delta(S_{1234xy})\delta(S_{123x45})
\end{equation}
As mentioned in Section~\ref{sec:connected-to-link}, the integrand is finite for a spurious solution of $S_{1234xy} = S_{123x45} = 0$, namely that with particles  1,2,3 and 4 collinear, as the ratio $\frac{(124)}{(123)}$ does not vanish.

We denote $S_1 \equiv S_{1234xy} $ and $S_2 \equiv S_{123x45}$. Consider first the GRT $\{f_1,f_2\}=0$ with $f_1 \equiv S_{1} $ and $f_2 \equiv  S_2 (123)(1xy)(3xy)(145)$.  The residue theorem then implies
\begin{align}
\label{eq:GRT16points}
\{S_1,S_2\} = - \{S_1,(123)\} - \{S_1,(1xy)\} - \{S_1,(3xy)\} - \{S_1,(145)\} = 0 
\eqndot
\end{align}
Note further that for $(123)=0$, $S_1$ factorises and thus
\begin{align}
\label{eq:GRT26points}
\{S_1,(123)\} = \{(234),(123)\}+\{(4xy),(123)\}+\{(y12),(123)\}+\{(3x1),(123)\} .
\end{align}
Plugging \eqref{eq:GRT26points} back into \eqref{eq:GRT16points}, we get
\begin{align}
\label{eq:GRT36points}
\begin{split}
&\textcolor{purple}{\{S_1,S_2\}} = - \textcolor{purple}{\{(234),(123)\}} - \{(4xy),(123)\} -\{(y12),(123)\}\\
& -\{(3x1),(123)\} - \{S_1,(1xy)\} - \{S_1,(3xy)\} - \{S_1,(145)\} = 0
\end{split}
\end{align}
Note the subtlety here: the two highlighted terms appear not to be distinct, since the configuration where $(123)=(234)=0$ is also a (spurious) solution of $S_1=S_2=0$. 
The fact that such a configuration appears after the application of the GRT 
follows from the requirement that the constraint $S_1=S_2=0$ in \eqref{eq:i5}
only includes non-spurious solutions, which for five points is not enforced
by the numerator.

Interestingly, the term $\{(234),(123)\}$ also highlights another phenomenon which does not occur for amplitudes.
For this term the integrand is given by
\begin{equation}
\frac{(13y)(23x)(124)(14x)}{(1xy)(3xy)(145)(4xy)(y12)\;S_2} \delta\big((234)\big)\delta\big((123)\big)
\eqncom
\end{equation}
and both the minor $(124)$ in the numerator as well as $S_{2}$ in the denominator
approach zero linearly if one parametrises the constraints imposed by the delta function. Under such a parametrisation, one finds that the direction in which the limit is taken changes the result. To calculate the correct residue, we have to take the limit ensuring that $ S_{1}$ is vanishing. We do so by setting $(123)=\varepsilon$ and
$(234)=\frac{(34x)(xy1)(24y)}{(4xy)(y12)(3x1)}\varepsilon$, and then letting $\varepsilon\to 0$.
Note that the term under consideration arises from factorizing $S_1$ in $\{S_1,(123)\}$; the limit ensures that $(123)=0$ is
approached precisely from the surface $S_1=0$.

The other residues coming from \eqref{eq:GRT36points} can be calculated straightforwardly.
Note that $S_1$ factorises for the terms $\{S_1,(1xy)\}$ and $\{S_1,(3xy)\}$; 
the resulting terms, together with those not involving $S_1$ in \eqref{eq:GRT36points}
are in one-to-one correspondence with the MHV$\times$MHV factorization poles
of the BCFW contour  \eqref{eq:FFcontour}. For the term $\{S_1,(145)\}$ one applies the GRT again, 
after which the calculation is identical to the four point case, and results
in all inverse soft contributions to the form factor.
For all terms, \eqref{eq:i5} gives the same residues as
the corresponding poles of the Graßmannian integral.

\subsection{Six points}

For the six point form factor, we checked numerically that the Graßmannian formula \eqref{eq:isfactor} evaluated on the conic constraints gives the correct result for the form factor. However, when attempting to perform a one-to-one mapping of the poles of this Graßmannian integral to those obtained from the BCFW contour (see Figure \ref{fig:contourgrid}) via the GRT, we find that it is impossible to identify all of them.
We furthermore collected evidence that even by using the identity \eqref{eq:id} repeatedly,
one might not be able to generate other representations which have all BCFW poles.

The six-point form-factor integrand in the inverse-soft-like representation \eqref{eq:isfactor} is given by
\begin{equation}
I_{6,3}=\frac{(13x)(13y)(23x)(124)(14x)(125)(135)(234)}{(123)(1xy)(3xy)(156)} \delta(S_{1234xy})\delta(S_{12345x}) \delta(S_{123456})
\eqncom
\label{eq:i6}
\end{equation}
and the poles contributing to the BCFW representation of the form factor can
be found in Figure~\ref{fig:contourgrid}.
Most of these poles can be recovered by successively applying the GRT to \eqref{eq:i6},
in particular all poles with $(156)=0$, corresponding to the inverse soft limit of $\ff_{5,3}$.

It is however impossible to find the poles $\{\pole{1},\pole{2},\pole{3}\}$
and $\{\pole{1},\pole{2},\poleu{5}\}$, corresponding to the factorization channels
$\amp_{6,2}\times\ff_{2,2}$ and $\amp_{5,2}\times\ff_{3,2}$. To see that these 
poles can never appear it is sufficient to realise that,
in the vicinity of these configurations, the integrand \eqref{eq:i6} is not singular enough to produce a finite residue.
Letting each of the vanishing minors at those poles approach zero as $\varepsilon\sim 0$, we find that for the respective configurations the integrand behaves as
\begin{equation}
\begin{aligned}
&\{\pole{1},\pole{2},\pole{3}\} \colon\quad
\frac{(124)(125)(135)(234)}{(123)\;S_{1234xy}S_{12345x}S_{123456}}\sim \frac{1}{\varepsilon^2}\eqncom\\
&\{\pole{1},\pole{2},\poleu{5}\}\colon\quad
\frac{(124)(234)}{(123)\;S_{1234xy}S_{12345x}S_{123456}}\sim \frac{1}{\varepsilon^2}\eqncom
\end{aligned}
\end{equation}
while in order for a residue to exist, the integrand would have to scale as $\varepsilon^{-3}$.
Since the GRT does not change this power counting, potential poles at these locations would be
cancelled by numerator factors. 

The identity \eqref{eq:id} can change the degree of divergence at configurations
away from the support of the conic constraints, i.e. at positions reached by the GRT.
In order to see if other representations of the integrand with the correct singularities at all BCFW poles
exist, we generated a very high number ($\O(10^6)$) of 
different representations of the integrand with a computer program, using the identity \eqref{eq:id} and cyclic symmetry,
and taking both \eqref{eq:i6} and \eqref{eq:Fn} as starting points.
We then checked that none of these representations has the correct degree of divergence at all BCFW poles.
This result is not conclusive, since we could only generate a finite number 
of representations due to computational constraints. In principle, the identity
\eqref{eq:id} can be applied over and over again. Nevertheless, our result is a very strong indication that there may not be any $G(3,8)$ representation based on the connected formula from which we can identify all BCFW terms one by one, although we emphasise once again that the connected formula does produce the correct form factor, namely the sum of all BCFW terms. 

Note however that some way of relating the formulations has to exist.
We speculate that it is possible to apply a GRT to \eqref{eq:i6},
and then to apply different identities \eqref{eq:id} to each of the resulting terms, effectively
combining different representations. Of course there is a proliferation of such possibilities without a clear physical motivation, and several attempts did not lead to the identification of the expected residues.
Since in any case the relation between the formulations is much more subtle 
compared to scattering amplitudes, it could be difficult to apply such a strategy systematically
to find the BCFW contour prescription in closed form beyond NMHV.
It remains to be investigated whether this tells us something about the physical properties of Graßmannian representations of (partially) off-shell observables. We leave this for future work.

\section{Conclusions}

In this note we investigated the contours of integration of the Graßmannian formulation of form factors proposed in \cite{Frassek:2015rka}, as well as the relation to the connected prescription for form factors \cite{Brandhuber:2016xue,He:2016jdg}. To this end, we used the on-shell diagram representation of form factors. The permutations labelling the bipartite on-shell diagrams allowed to obtain the linear relations among the minors in the Graßmannian formula for a given diagram, and thus deduce the corresponding contour of integration. We applied this procedure explicitly to NMHV form factors, arriving at a compact form of the contour given in \eqref{eq:FFcontour}, which is the analogue of the odd-even form of the NMHV contour for amplitudes \cite{ArkaniHamed:2009dg}. As we emphasised, this method should apply to general form factors beyond the NMHV case. It would be of interest to investigate similarities and differences between the contours of integration for general N$^{k-2}$MHV amplitudes and form factors.

We then studied the connected prescription for form factors,
lifting this formulation to the Graßmannian. In particular we provided a representation of this Graßmannian formula which has the same recursive structure as its amplitude
counterpart. In this representation each additional particle is added via a factor which ensures the correct behaviour in the soft limit.
Analysing this formulation using the global residue theorem, we were able to show that the connected prescription also non-trivially gives rise to the BCFW contour obtained from on-shell diagrams for four and five points.  We found that a new feature arises already at five points, where a $0/0$ term appears. This requires a careful treatment, in particular regarding the direction in which the pole is approached. At six points, we first checked that the connected prescription formula gives the same results as the BCFW formula. Interestingly, we also found strong evidence that through a direct application of GRTs it may not be possible to perform a one-to-one mapping between the poles present in the connected prescription and in the BCFW contour. This situation is quite different from that of on-shell scattering amplitudes, for which the two formulas can be smoothly deformed into one another, and it may teach us important lessons about applying the Graßmannian formalism and the connected prescription to form factors or more general off-shell and/or non-planar objects. As a way forward it may be fruitful to note the role such smooth deformations play in showing the equivalence of similar integral formulas in the case of form factors of Wilson line operators \cite{Bork:2017qyh}.

Form factors provide a bridge between on-shell scattering amplitudes and completely off-shell correlation functions, and thus they are ideal objects for a better understanding of how the Graßmannian integral and on-shell diagrams can be generalised to off-shell quantities. 
The recent progress in studying correlation functions in terms of amplituhedron-like geometries \cite{Eden:2017fow} raises hope that these methods are indeed more generally applicable for a variety of observables in $\N=4$ SYM. It would be interesting
to see if form factors can interpolate between the geometries corresponding to amplitudes and correlation functions.
Furthermore, form factors are intrinsically non-planar, even in the large-$N$ limit, which may be one of the main causes of the new features we found in the study of the connected prescription using the GRT. It would therefore be interesting to explore applications of the recent developments concerning non-planar on-shell diagrams \cite{Franco:2015rma, Bourjaily:2016mnp} to form factors. Finally, it would be interesting to explore the interplay between ambitwistor strings and on-shell diagrams, studied in \cite{Farrow:2017eol} for amplitudes in $\N=4$ SYM and $\N=8$ supergravity, for form factors at loop level.

\section*{Acknowledgements}
We would like to thank
Andi Brandhuber,
Ed Hughes,
Rodolfo Panerai, 
Gregor Richter,
Matthias Staudacher,
Gabriele Travaglini
 and especially 
Matthias Wilhelm
for very interesting discussions.
We thank L.V. Bork and A.I. Onishchenko for clarifying various points regarding
their recent work \cite{Bork:2017qyh}.
DM received support from GK 1504 \emph{``Masse, Spektrum, Symmetrie''}.
The work of BP was supported by the ERC starting grant 637019 ``\emph{MathAm}''. The work of CW was supported in part by a DOE Early Career Award under Grant No. DE-SC0010255.  DN is supported by the STFC consolidated grant “Particle Physics at the Higgs Centre”, by the National Science Foundation. BP would like to thank Humboldt University of Berlin, where part of this work was accomplished. DN and CW would like to thank the support and hospitality of the KITP program \emph{``Scattering Amplitudes and Beyond''} at UCSB, where last stages of this work were carried out. DN and CW's research was supported in part by the NSF under Grant No. NSF PHY-1125915. DN would also like to thank Walter Burke ITP at Caltech and QMAP at UC Davis for hospitality during final stages of this work.

\bibliographystyle{utphys2}
\bibliography{refs}

\end{document}